\documentclass[aps,prb,twocolumn,superscriptaddress,showpacs,floatfix]{revtex4-1}
\usepackage{color}
\usepackage[caption=false]{subfig}
\usepackage{bm}        % for math
\usepackage{amssymb}   % for math
\usepackage{amsmath}
\usepackage{gensymb}   % for \degree symbol
\usepackage{bbm}
\usepackage{hyperref}
\usepackage{graphicx}
    \usepackage[caption=false]{subfig}
\usepackage{array}
\usepackage{braket}
\usepackage{natbib}
\usepackage [latin1]{inputenc}
\usepackage{siunitx}

\newcommand{\bea}{\begin{eqnarray}}
\newcommand{\eea}{\end{eqnarray}}

\newcommand{\br}{\mathbf{r}}

\newcommand{\be}{\begin{equation}}
\newcommand{\ee}{\end{equation}}

\newcommand{\beal}{\begin{align}}
\newcommand{\eeal}{\end{align}}
\newcommand{\ra}{\rangle}
\newcommand{\la}{\langle}

\newcommand{\upa}{\uparrow}
\newcommand{\dna}{\downarrow}

\newcommand{\dg}{{\dagger}}
\newcommand{\pdg}{{\phantom\dagger}}

\begin{document}

\title{Multipolar magnetism in d-orbital systems: Crystal field levels, octupolar order, and orbital loop currents}
\author{Sreekar Voleti}
\affiliation{Department of Physics, University of Toronto, 60 St. George Street, Toronto, ON, M5S 1A7 Canada}
\author{D. D. Maharaj}
\affiliation{Department of Physics and Astronomy, McMaster University, Hamilton, ON L8S 4M1 Canada}
\author{B. D. Gaulin}
\affiliation{Department of Physics and Astronomy, McMaster University, Hamilton, ON L8S 4M1 Canada}
\affiliation{Brockhouse Institute for Materials Research, McMaster University, Hamilton, ON L8S 4M1 Canada}
\affiliation{Canadian Institute for Advanced Research, 661 University Ave., Toronto, ON M5G 1M1 Canada}
\author{Graeme Luke}
\affiliation{Department of Physics and Astronomy, McMaster University, Hamilton, ON L8S 4M1 Canada}
\affiliation{Brockhouse Institute for Materials Research, McMaster University, Hamilton, ON L8S 4M1 Canada}
\affiliation{TRIUMF, 4004 Wesbrook Mall, Vancouver, BC, V6T 2A3, Canada}
\author{A. Paramekanti}
\email{arunp@physics.utoronto.ca}
\affiliation{Department of Physics, University of Toronto, 60 St. George Street, Toronto, ON, M5S 1A7 Canada}
\date{\today}
\begin{abstract}
Quantum magnets with spin $J=2$, which arise in spin-orbit coupled Mott insulators,
can potentially display multipolar orders.
Motivated by gaining a better microscopic understanding of the local physics of such $d$-orbital
quantum magnets, we carry out an exact diagonalization study of a simple
octahedral crystal field Hamiltonian for two electrons, incorporating spin-orbit coupling (SOC) and interactions. 
While the rotationally invariant Kanamori interaction in the $t_{2g}$ sector leads to a five-fold degenerate $J=2$ manifold, we 
find that either explicitly including the $e_g$ orbitals, or going beyond the rotationally
invariant Coulomb interaction within the $t_{2g}$ sector, causes a degeneracy breaking of the $J\!=\!2$ levels. 
This can lead to a low-lying
non-Kramers doublet carrying quadrupolar and octupolar moments and an excited triplet which supports magnetic dipole moments,
bolstering our previous phenomenological proposal for the
stabilization of ferro-octupolar order in heavy transition metal oxides. We show that the spontaneous time-reversal symmetry breaking due to 
ferro-octupolar ordering within the non-Kramers doublet leads to electronic orbital loop currents. The resulting internal
magnetic fields can potentially explain the small fields inferred from muon-spin relaxation
($\mu$SR) experiments on cubic $5d^2$ osmate double perovskites Ba$_2$ZnOsO$_6$, Ba$_2$CaOsO$_6$, and Ba$_2$MgOsO$_6$,
which were previously attributed to weak dipolar magnetism. We make further predictions for oxygen NMR experiments on these materials. We
also study the reversed level scheme, where the $J\!=\!2$ multiplet splits into a low-lying magnetic triplet and excited non-Kramers doublet, 
presenting single-ion results for the magnetic susceptibility in this case, and pointing out its possible relevance 
for the rhenate  Ba$_2$YReO$_6$. Our work highlights the intimate connection between the physics of heavy transition metal oxides and
that of $f$-electron based heavy fermion compounds.
\end{abstract}
\maketitle

Multipolar orders have been proposed and discussed extensively in $f$-orbital based heavy fermion compounds \cite{MultipolarRMP2009,HauleKotliar2009,SantiniNpO2_PRL2000,NpO2TripleQ_PRL2002,Fazekas_PRB2003, NpO2NMR_PRL2006,Arima2013,Sakai_JPSJ2011,Sato_PRB2012,Nakatsuji_PRL2014,Hattori2016,Freyer2018,SBLee2018,Patri2019}.
Such multipolar orders have also been proposed to occur in $d$-orbital metals with large spin-orbit coupling (SOC),
such as LiOsO$_3$ and Cd$_2$Re$_2$O$_7$, via Pomeranchuk instabilities of the Fermi liquid \cite{LFu_PRL2015}.
Optical second-harmonic generation experiments on Cd$_2$Re$_2$O$_7$ have found evidence for such an inversion broken
quadrupolar ordered state below  $T_c \!\sim\! 200$\,K
\cite{Hsieh_Science2017}. Other candidates for multipolar orders include proposed quadrupolar order in
A$_2$OsO$_4$ (with A = K,Rb,Cs) \cite{Motome2018}.

In recent work, we have studied $d$-orbital Mott insulators with large SOC and a $d^2$ configuration in a local octahedral environment, and
proposed these systems as candidates for realizing ferro-octupolar order \cite{maharaj2019octupolar,paramekanti2019octupolar}. 
Previous studies of such $d^2$ quantum magnets \cite{ChenBalents2010,ChenBalents2011,Svoboda_PRB2017} have argued that the 
combination of crystal field and interaction effects, leads to the stabilization
of a state with total $L\!=\! 1$ and $S\!=\! 1$, which are locked by SOC into a $J\!=\!2$ spin. Motivated by experiments 
\cite{Thompson_JPCM2014,Kermarrec2015,Thompson_PRB2016,MarjerrisonPRB2016,maharaj2019octupolar} on certain cubic
double perovskite (DP) Mott insulators, Ba$_2$ZnOsO$_6$, Ba$_2$CaOsO$_6$, and Ba$_2$MgOsO$_6$, which host a $5d^2$ configuration
on Os, we have instead proposed \cite{paramekanti2019octupolar}
that their observed nontrivial phenomenology, such as entropy and a spin gap, 
could be captured by assuming that the five-fold $J\!=\! 2$ multiplet is weakly split, resulting in a ground state
non-Kramers doublet carrying quadrupolar and octupolar moments.
The lack of any observed crystal distortions in X-ray and neutron diffraction experiments appears to rule out quadrupolar order \cite{maharaj2019octupolar}.
Uniform ferro-octupolar ordering in the low lying doublet manifold then
provides the most viable route to further reconciling the cubic symmetry, the observation 
of time-reversal symmetry breaking seen via $\mu$SR oscillations \cite{Thompson_JPCM2014},
the apparent lack of any magnetic Bragg peaks in elastic neutron diffraction experiments \cite{maharaj2019octupolar},
and the spin gap observed in inelastic neutron scattering experiments \cite{maharaj2019octupolar,paramekanti2019octupolar}.

In this paper, we provide further theoretical calculations in favor of the above scenario. We first present exact diagonalization results on a simple local crystal field
Hamiltonian keeping the $t_{2g}$ and $e_g$ levels in an octahedral environment, showing that the combination of SOC and interactions does
favor a non-Kramers ground state doublet. We show how the splitting between this doublet and the excited magnetic triplet depends on SOC and
the Hund's coupling and results from perturbative $t_{2g}$-$e_g$ mixing. Such $t_{2g}$-$e_g$ mixing was discussed previously but its
importance for the low energy physics appears not to have been properly recognized \cite{ChenBalents2011,Fiete_eg_PRB2018}.
We also examine a model of just $t_{2g}$ electronic states, and show that deviations of the Coulomb interaction from spherical symmetry,
perhaps engendered by hybridization with oxygen orbitals \cite{Ribic_PRB2014}, can
lead to a similar non-Kramers doublet state. This doublet-triplet splitting may be too small to be resolved using
resonant inelastic X-ray scattering experiments \cite{BoYuan_PRB2017,Paramekanti_PRB2018}, 
but it is crucial for the low energy symmetry-breaking orders.
We study the impact of ferro-octupolar order within this low energy non-Kramers doublet,
and show that this leads to orbital electronic currents, generating internal magnetic fields and
semi-quantitatively explain the $\mu$SR oscillations seen in Ba$_2$ZnOsO$_6$, Ba$_2$CaOsO$_6$, and Ba$_2$MgOsO$_6$.
The non-spherical Coulomb interaction mechanism for splitting the $J\!=\! 2$ multiplet discussed above also permits for the possibility for the 
level ordering to be reversed, with a magnetic triplet ground state and an excited non-Kramers doublet.
We present single ion results for the magnetic susceptibility in this case, arguing that this 
reversed level scheme is likely to be relevant to the $5d^2$ rhenate  \cite{Aharen_PRB2010} Ba$_2$YReO$_6$.

Our theory strengthens the case for multipolar orders in a class of $d$-orbital Mott insulators, pointing to a smooth
conceptual link between the physics of heavy $d$-orbital oxides and $f$-electron based heavy fermion materials.
Such octupolar order with a high transition temperature may provide a new template to store information.

\section{Local model}
We use the following Hamiltonian for two electrons in a $d$-orbital placed in an octahedral environment:
\bea
H=H_{\rm CEF}+H_{\rm SOC}+H_{\rm int}
\label{eq:hfull}
\eea
where we include the octahedral crystal field splitting, SOC, and Kanamori interactions, written in the orbital basis 
($\{yz,xz,xy\},\{x^{2}\!-\!y^{2},3z^{2}\!-\!r^{2}\} ) \leftrightarrow (\{1,2,3\},\{4,5\}$) where $\alpha\equiv \{1,2,3\}$ label $t_{2g}$
orbitals and $\alpha\equiv \{4,5\}$ label $e_g$ orbitals.
The CEF term is given by:
\begin{equation} \label{cfham}
H_{\rm CEF}=V_C\sum_{\alpha=4,5}\sum_{s}n_{\alpha,s}
\end{equation}
where $s$ is the spin. The SOC term is
\begin{align}
\begin{split}
H_{\rm SOC} &= {\lambda \over 2} \sum_{\alpha, \beta} \sum_{s,s'} \bra{\alpha}\mathbf{L}\ket{\beta} \cdot \bra{s}\pmb{\sigma}\ket{s'}c^\dagger_{\alpha,s} c_{\beta, s'}
\end{split}
\end{align}
where $\pmb{\sigma}$ refers to the vector of Pauli matrices, and $\mathbf{L}$ is the orbital angular momentum. Its components in the orbital basis are 
given in Appendix A. We assume a Kanamori interaction for all five $d$-orbitals given by
\bea
\!\!\!\! H_{\rm int} &=& U\sum_{\alpha}n_{\alpha \uparrow}n_{\alpha \downarrow} \!+\! U' \sum_{\alpha > \beta} n_\alpha n_\beta 
\!-\! J_H \sum_{\alpha \neq \beta} \vec S_\alpha \cdot \vec S_\beta \nonumber \\
&+& J_H \sum_{\alpha\neq\beta} c^\dg_{\alpha \upa} c^\dg_{\alpha\dna} c^\pdg_{\beta \downarrow} c^\pdg_{\beta \upa}
\eea
where $\vec S_\alpha = (1/2) c^\dg_{\alpha s} \vec \sigma_{s,s'} c^\pdg_{\alpha s'}$. This simple form, where we use the same interaction parameters 
for all $t_{2g}$ and $e_g$ orbitals, is used to avoid a proliferation of interaction parameters. Assuming spherical symmetry of the 
Coulomb interaction, we have $U' = U - 2 J_H$ (see, for e.g., Ref.\onlinecite{Georges2013}).

For electronic configurations with partially filled $t_{2g}$ orbitals,
the most commonly used approach is to simply ignore the $e_g$ orbitals and
restrict attention to the low energy $t_{2g}$ states. We find that the ground state manifold in this approximation consists
of a five-fold degenerate $J\!=\! 2$ state.
However, we show below that this degeneracy is
further split due to two possible microscopic mechanisms: $t_{2g}$-$e_g$ mixing and deviations of the Coulomb
interaction from spherical symmetry.

\subsection{$t_{2g}$-$e_g$ mixing: Exact results, perturbation theory}

\begin{figure}[t]
	\includegraphics[width = 0.49\textwidth]{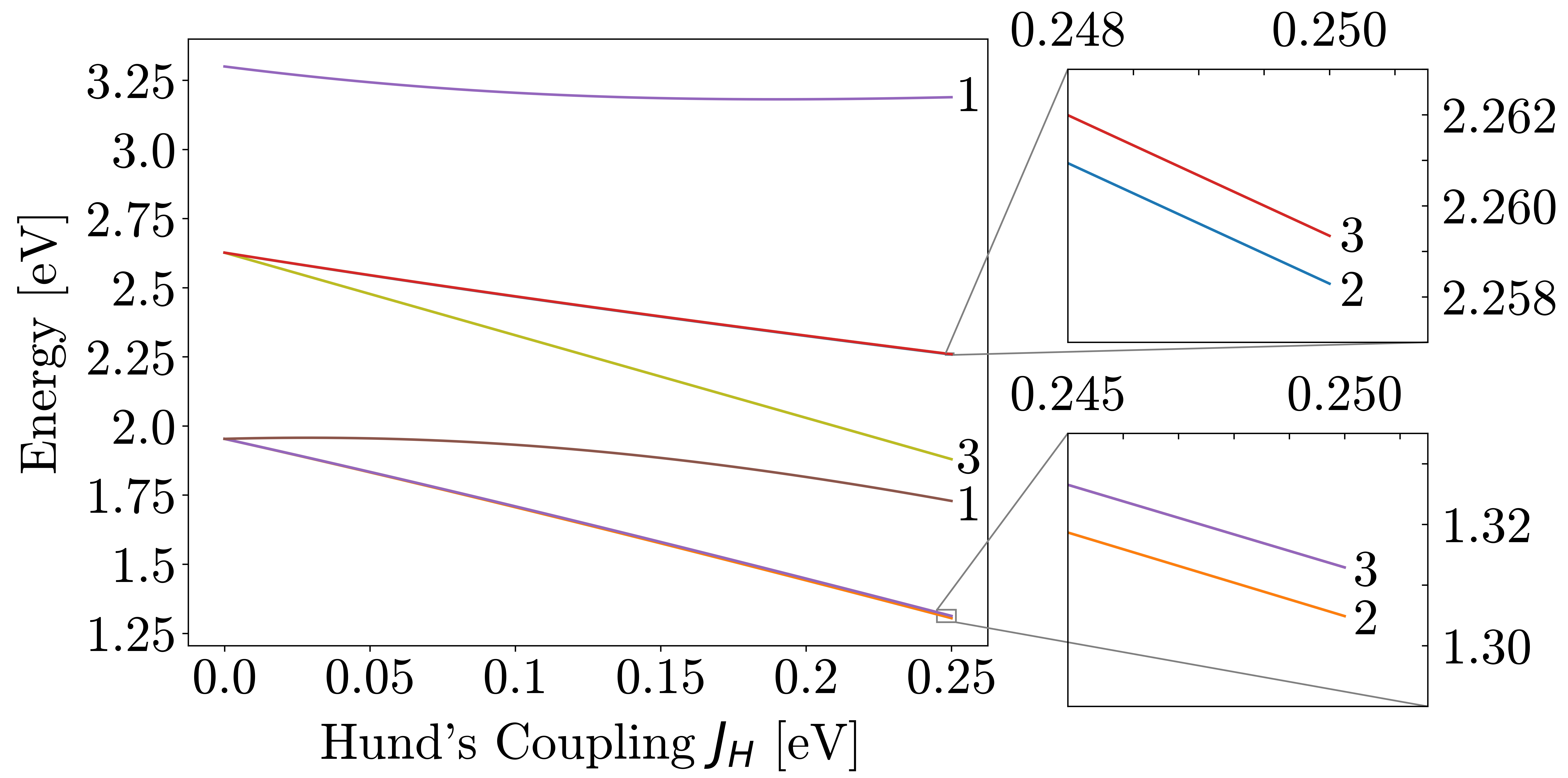}
	\caption{Low energy spectrum ($15$ lowest eigenvalues) of the Hamiltonian in Eq.~\ref{eq:hfull} with two electrons, corresponding to states where both electrons 
	predominantly occupy the $t_{2g}$ orbitals. The numbers at the end of the curves, and in the zoomed-in insets which show weak splittings, 
	indicate the degeneracies of the different energy levels.}
	\label{fig:lamdep}
\end{figure}

We consider two electrons in the full $d$-orbital manifold including
$t_{2g}$ and $e_g$ states, and study this using numerical exact diagonalization in the $45$ basis states.
For coupling constants, we use values typical for $5d$ transition metal oxides: 
$V_C \! = \! 3$\,eV, $U \! = \! 2.5$\,eV, $\lambda \! = \! 0.4$\,eV, 
and $J_H \! = \! 0.25$\,eV.
Fig.\ref{fig:lamdep} plots the evolution with $J_H$ of the lowest $15$ energy levels which correspond to 
eigenstates where the two electrons are predominantly both in the $t_{2g}$ sector. The indicated numbers mark the
degeneracies of these multiplets. For $J_H\!=\!0$, there are just three energy levels, which, in increasing order of energy, 
correspond to having (i) both electrons in $j\!=\!1/2$, (ii) one electron in $j\!=\!1/2$ and one electron in $j\!=\!3/2$
(energy cost $3\lambda/2$), and (iii) both electrons in $j\!=\!3/2$ (energy cost $3\lambda$). We see that the lowest
energy set of $5$ states evolves adiabatically out of the first sector as we increase $J_H$; this set of five states
corresponds to the $J\!=\!2$ moment. However, a zoom-in of this multiplet, as well as of one of the higher energy 
multiplets, shows that the apparent five-fold degeneracy
of these states is actually weakly broken as $2\oplus 3$ due to weak $t_{2g}$-$e_g$ mixing. 
In particular, the naively expected five-fold degenerate 
$J\!=\! 2$ ground state is 
split into a non-Kramers doublet ground state and an excited magnetic triplet; for the typical values listed above, 
this splitting is $\sim\! 8$\,meV. 

\begin{figure}[t]
  \includegraphics[width=0.49\textwidth]{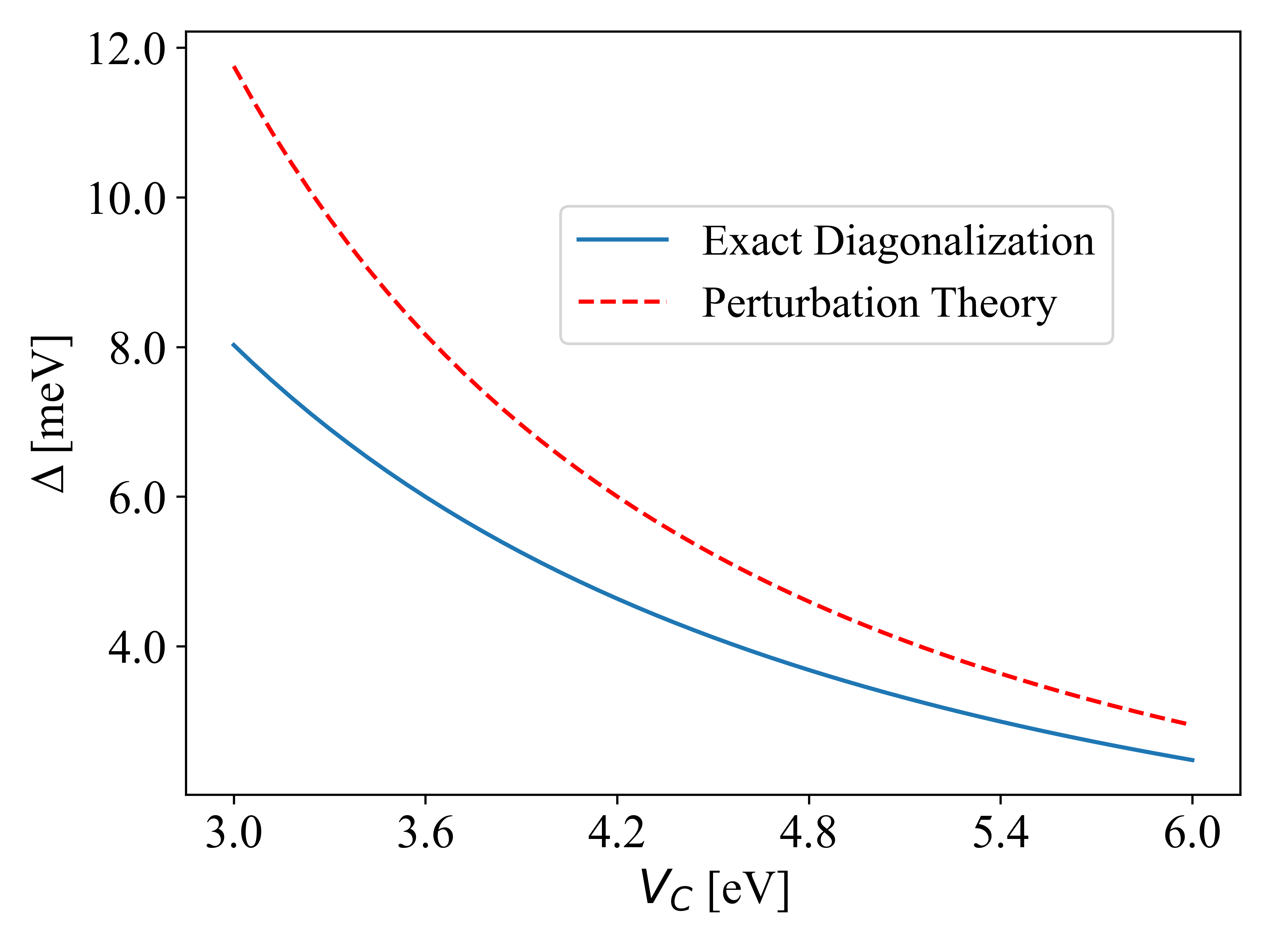}
  \caption{Energy difference between the lower energy non-Kramers doublet ($E_d$) and the excited triplet ($E_t$), given by $\Delta = E_t-E_d$,
  obtained via exact diagonalization of the Hamiltonian in Eq.\ref{eq:hfull} (blue, solid)
  plotted as a function of the dominant $t_{2g}$-$e_g$ splitting $V_C$. We compare this with
  the third order perturbation theory result (red, dashed) induced by small $(J_H/V_C,\lambda/V_C)$ 
  which leads to weak $t_{2g}$-$e_g$ mixing.}
  \label{fig:Deldep}
\end{figure}

Fig.~\ref{fig:Deldep} shows the dependence of this lowest energy doublet-triplet energy splitting (blue solid line) on $V_C$.
We find that this splitting can be semi-quantitatively captured within third order perturbation theory, as discussed in Appendix B,
where we first eliminate the $e_{g}$ states, to find an effective $t_{2g}$ model, and then diagonalize this reduced Hamiltonian. The relevant
terms arise at ${\cal O}(\lambda^2 J_H/V_C^2)$,
from the following sequence: (i) SOC $\lambda$ promoting one electron from the $t_{2g}$ manifold into the $e_{g}$ sector, (ii) intermediate state
$t_{2g}$-$e_g$ interactions driven by Hund's coupling set by $J_H$, and finally (iii) de-exciting back via SOC $\lambda$ to end up with
both electrons in the $t_{2g}$ manifold. Diagonalizing this third-order perturbative Hamiltonian, in conjunction with the bare $t_{2g}$ Hund's coupling,
leads to the non-negligible splitting shown (red dashed line) in Fig.~\ref{fig:Deldep}, which
agrees well with the full numerical calculation in the regime of large $V_C$. Our result is in
contrast with a previous conjecture that the splitting would appear at fourth-order in 
perturbation theory \cite{ChenBalents2011}, which would have indeed rendered this effect negligible.
This highlights a non-trivial effect of $t_{2g}$-$e_g$ mixing, showing that it can be important
for nucleating multipolar order in $5d$ Mott insulators. However, this effect by itself may be too small to account
for the spin gap observed in neutron scattering experiments \cite{maharaj2019octupolar,paramekanti2019octupolar} on
Ba$_2$ZnOsO$_6$, Ba$_2$CaOsO$_6$, and Ba$_2$MgOsO$_6$. We next turn to an 
additional mechanism, which can cooperate to enhance this splitting, or even reverse the level ordering which
we argue is important in certain other materials.

\subsection{Non-spherical Coulomb interactions in $t_{2g}$ model}

\begin{figure}[t]
  \includegraphics[width=\linewidth]{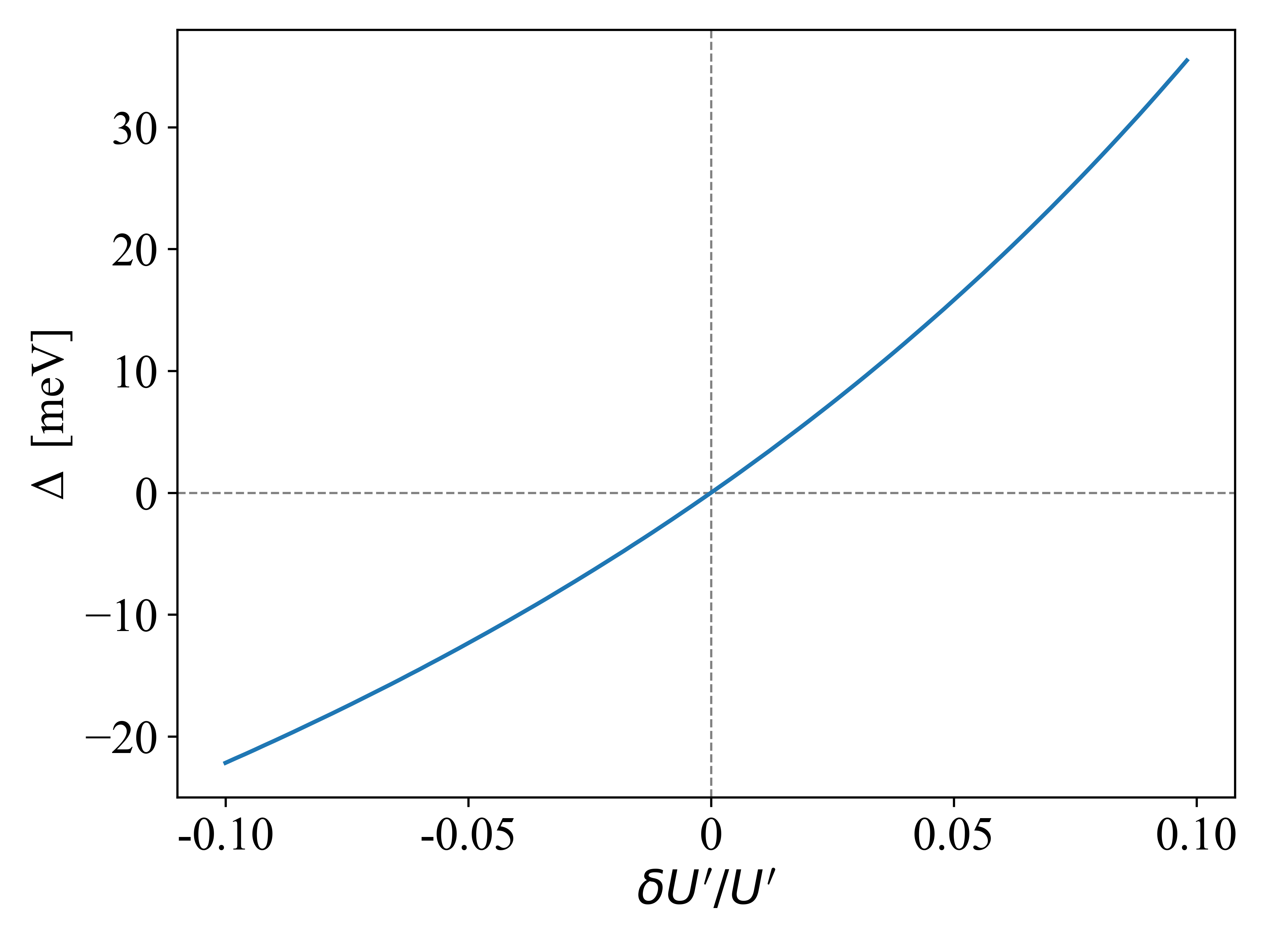}
  \caption{Energy difference $\Delta = E_t - E_d$ between the magnetic triplet and the non-Kramers doublet obtained via
  exact diagonalization of the $t_{2g}$-only model, shown as a function of the normalized deviation $\delta U' / U'$ of the Coulomb interaction
  from spherical symmetry. For $\delta U' > 0$, the non-Kramers doublet has lower energy so $\Delta > 0$.}
  \label{fig:gapdeltau}
\end{figure}

The second important physical effect we consider is that projecting  the Coulomb interaction to the $t_{2g}$ Wannier orbitals 
can lead to deviations from the spherical symmetry assumption, so that $U' \neq U - 2 J_H$. This is expected to be more
important for $5d$ orbitals which have more significant overlap with the oxygen cage, as has been previously
noted in an {\it ab initio} study \cite{Ribic_PRB2014}.
We therefore numerically diagonalize the above model Hamiltonian, restricting ourselves to the Hilbert space where both
electrons occupy the $t_{2g}$ orbitals, and varying
$\delta U' \!=\! U' \!-\! (U \!-\! 2J_H)$ to simulate the deviation from spherical symmetry. 
Fig.\ref{fig:gapdeltau} shows how the low energy degeneracy gets split as we go away from
$\delta U' \!=\! 0$. We
see from here that even a small deviation $\delta U'/U' \!\sim\! 0.1$ leads to a substantial splitting $\sim\! 20$\,meV. For
$\delta U' \!>\! 0$, we find that the non-Kramers doublet is lower in energy than the magnetic triplet, which we argue is
relevant to osmates such as Ba$_2$ZnOsO$_6$, Ba$_2$CaOsO$_6$, and Ba$_2$MgOsO$_6$. The case where the
$\delta U' \!<\! 0$, so that the magnetic triplet lies lower in energy than the doublet, 
may be important to understand aspects of the unusual magnetism of the rhenate \cite{Aharen_PRB2010} Ba$_2$YReO$_6$; this will be discussed
in Section III.

\section{Magnetic fields from octupolar order}
On phenomenological grounds, and the above microscopic calculations, $5d^2$ oxides are candidates for a low-lying non-Kramers doublet. As shown 
previously \cite{paramekanti2019octupolar}, 
this doublet may be described using the wavefunctions of the $J\!=\!2$ manifold in terms of $|J_z\ra$ eigenstates written as pseudospin-$1/2$ states:
\bea
|\psi_{g,\uparrow}\rangle = |0\rangle;~~~
|\psi_{g,\downarrow}\rangle = \frac{1}{\sqrt{2}} (|2\rangle + | -2 \rangle)
\eea
Each of these two states is individually time-reversal invariant. 
The angular momentum operators $(J_x^2-J_y^2)$ and $(3 J_z^2-J^2)$, restricted to this basis, act as
pseudospin-$1/2$ operators $(\tau^{x}, \tau^z)$, forming the two
components of an XY-like quadrupolar order parameter, while $\overline{J_x J_y J_z}$ (with overline denoting symmetrization) behaves
as $\tau^y$, and serves as the Ising-like octupolar order parameter.
The mean field ferro-octupolar ordered ground state is described by each site being in the superposition state $|\psi^{\rm oct}_\pm\ra = |\psi_{g,\upa}\ra \pm i |\psi_{g,\dna}\ra$.
Here, the signs reflect the $Z_2$ nature of the Ising order, and `$i$' reflects the breaking of time-reversal symmetry. 

The broken time-reversal symmetry of the octupolar ground state would lead to internal magnetic fields in the crystal. Using 
exact diagonalization, we obtain $|\psi^{\rm oct}_\pm\ra$ as the two-electron wavefunction obtained by superposing the two degenerate time-reversal invariant 
ground eigenstates as above, and compute the electronic currents in these states which generate the internal magnetic fields.
In the single-site picture, the orbital currents responsible for the internal fields
live on the $d^2$ ion. We thus define the orbital current density operator as 
\begin{align}
\mathbf{J}(\br) = {ie\hbar \over 2m} \sum_s  \left( \Psi_s^\dagger (\pmb{\nabla}\Psi^\pdg_s) - (\pmb{\nabla} \Psi^\dagger_s) \Psi^\pdg_s \right)
\end{align}
where $s$ sums over the physical electron spin. We expand the operator $\Psi$ in the orbital basis as
\begin{equation}
\Psi_s^\dagger = \sum_{\alpha} \psi_{n\ell \alpha}(r,\theta,\phi) c^\dg_{\alpha,s}
\end{equation}
where $\br \equiv (r,\theta,\phi)$, $\psi_{n\ell\alpha}$ refers to the real hydrogen-like wavefunction, with $n=5$ and $\ell=2$ for the $5d$ wavefunctions, and
$\alpha$ denotes the orbital. We thus
arrive at the spatially varying expectation value of the current density operator:
\bea
\la \mathbf{J}(\br) \ra_{\pm} 
&=& {ie\hbar \over 2m} \sum_s \sum_{\alpha\beta} \la \psi^{\rm oct}_\pm | c^\dagger_{\alpha,s}c^\pdg_{\beta,s}| \psi^{\rm oct}_\pm\ra~ \pmb{\xi}_{\alpha \beta} \\
\pmb{\xi}_{\alpha \beta} &=& R_{n \ell }^2(r) \left( Y_{\ell \alpha} \pmb{\nabla} Y_{\ell \beta} - Y_{\ell \beta} \pmb{\nabla} Y_{\ell \alpha} \right)
\eea
where the two Ising states have $\la \mathbf{J}(\br) \ra_- = - \la \mathbf{J}(\br) \ra_+$.
Here, $Y_{\ell \alpha}(\theta,\phi)$ are real Tesseral harmonics, and $R_{n\ell}(r)$ is the radial wavefunction. 
To compute the current density, we use a variational ansatz for the radial
wavefunction, which takes on a hydrogenic form, but with an effective nuclear charge which decreases with $r$, from a 
bare nuclear charge $Z_0$ for $r \!\to\! 0$ to the screened effective charge $Z_{\infty}$ for $r\!\to\!\infty$, over a length scale $r_0$.
For the Os$^{6+}$ ion relevant to Ba$_2$ZnOsO$_6$, Ba$_2$CaOsO$_6$, and Ba$_2$MgOsO$_6$, we use $Z_0\!=\!76$ and $Z_\infty\!=\!7$,
and consider different values of $r_0$; details are given in Appendix B.

\begin{figure} [t]
        \includegraphics[width=0.23\textwidth]{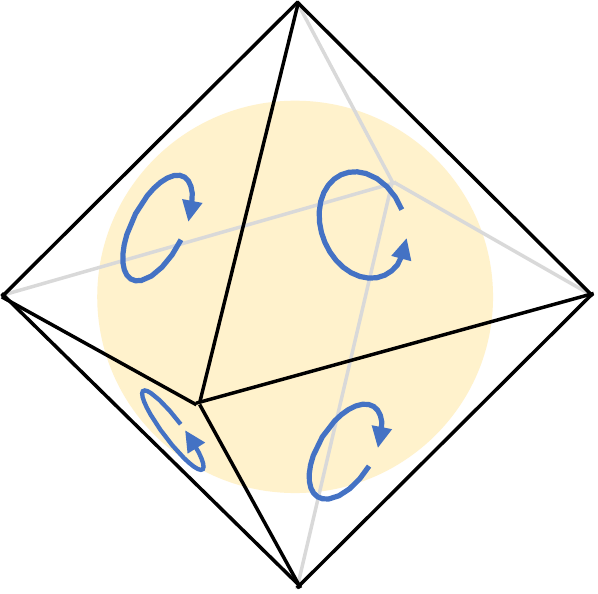}
        \includegraphics[width=0.23\textwidth]{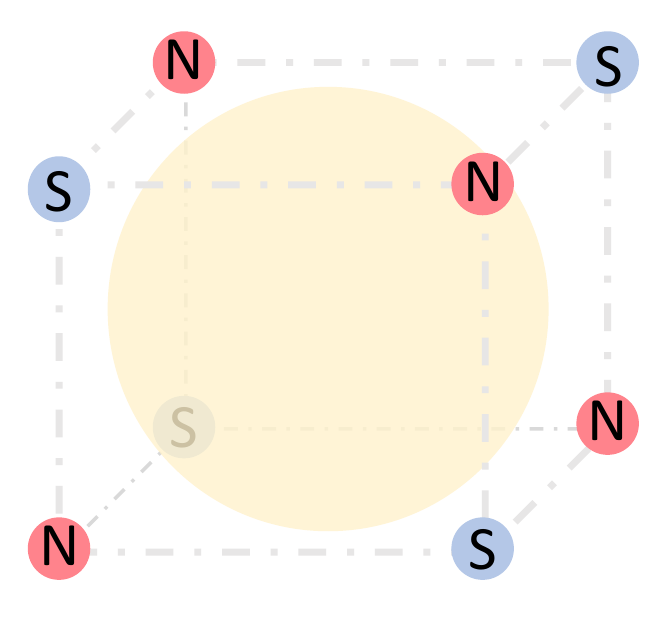}
    \caption{Left: Schematic plot of the orbital current pattern on the $5d^2$ Os ion (indicated by the ball), showing that it has the same symmetry as plaquette 
    loop current order residing on the OsO$_6$ octahedral cage. Right: Configuration of fictitious ``magnetic monopoles'' forming an octupole,
    which would produce the octupolar current loop pattern shown in the left panel.}
           \label{fig:octupole}
\end{figure}

Using this expectation value for the current density, we compute the magnetic field via
\begin{equation} \label{mfield}
\mathbf{B}_\pm(\mathbf{r}) = {\mu_0 \over 4\pi} \int d^3 r' \text{ }{ \braket{\mathbf{J(r')}}_\pm \times (\mathbf{r-r'}) \over |\mathbf{r-r'}|^3 }
\end{equation}
where the integral is carried out over primed variables. The two Ising time-reversed partner states
have opposite magnetic fields ${\bf B}_{-}(\br) = - {\bf B}_{+}(\br)$.

\begin{figure}[t]
  \includegraphics[width=0.5\textwidth]{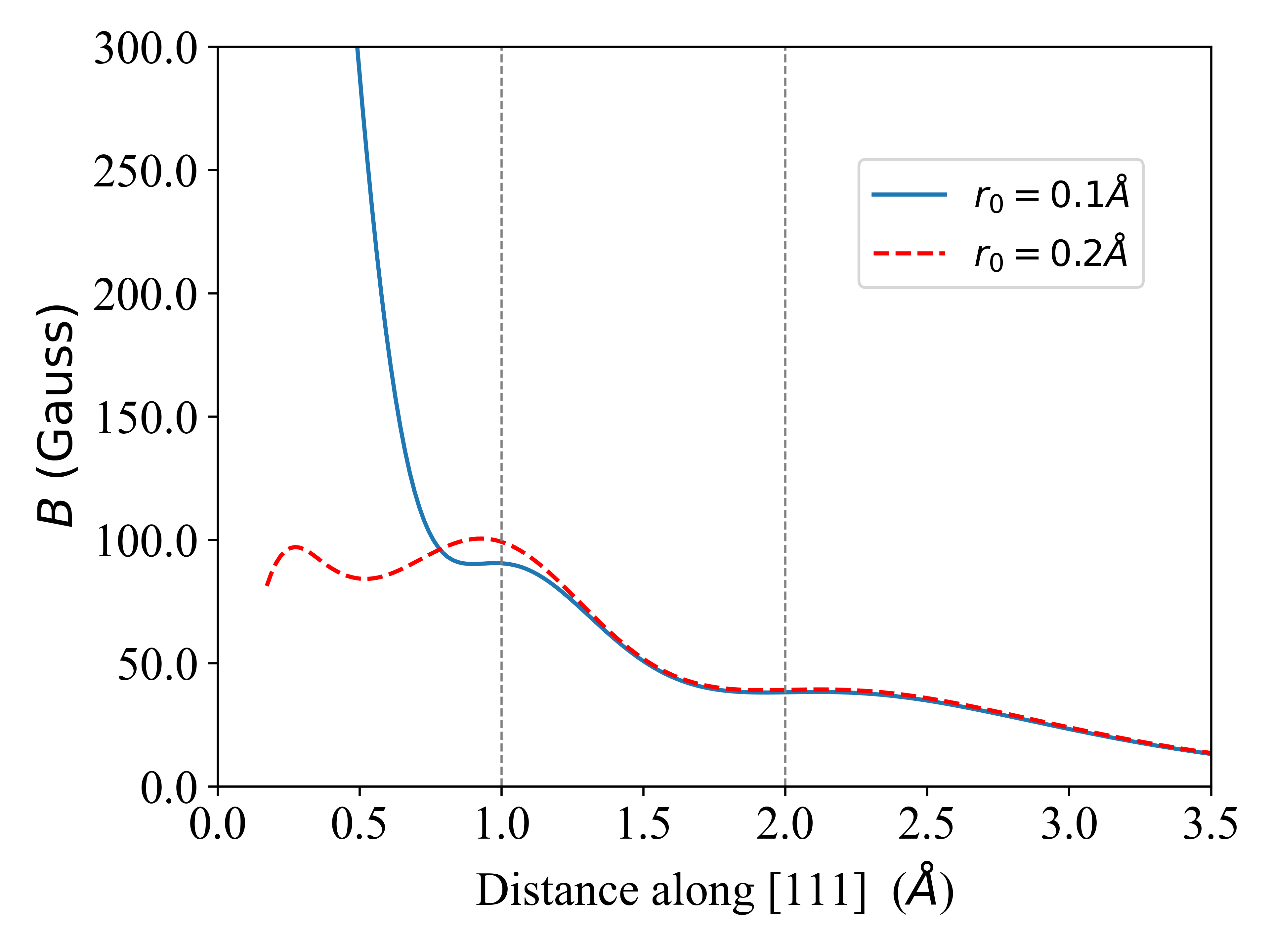}
  \caption{Magnetic field generated within the crystal in the presence of ferro-octupolar order, plotted as a function of distance from the $5d^2$ Os ion along the $[111]$ direction.
  The two curves correspond to different choices of the screening parameter $r_0$, which impacts the field only at short distances. The wiggles reflect the structure of the radial wavefunction.}  
 \label{fig:magfield}
\end{figure}

\begin{figure}[b]
  \includegraphics[width=0.5\textwidth]{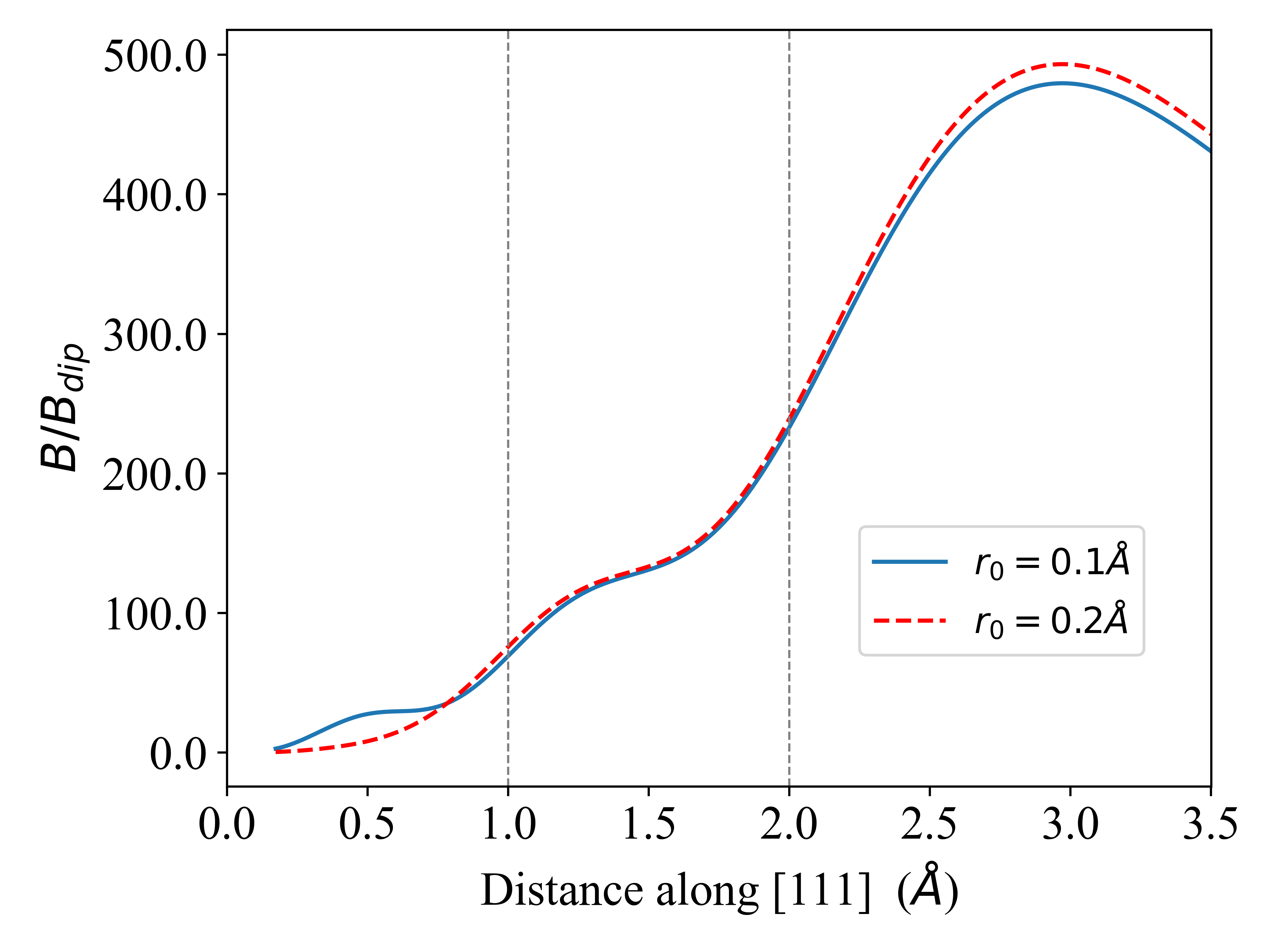}
  \caption{Magnetic field in the presence of ferro-octupolar order, plotted as a function of distance from the $5d^2$ Os 
  ion along the $[111]$ direction.
  The data are the same as in Fig.~\ref{fig:magfield}, but normalized by $B_{dip}$ which denotes the magnetic field at the same location
  generated by a $1 \mu_B$ dipole moment located at the origin and pointing along the $[111]$ direction.}
  \label{fig:magratio}
\end{figure}

The orbital current pattern which creates this field is shown schematically in Fig.~4 (left panel), highlighting that it
is analogous to loop current orders proposed in certain cuprate and heavy fermion materials \cite{Varma_PRL2002,Chandra_Nat2002}. 
In a more realistic calculation, which retains hybridization
with oxygen, the octupolar order we have uncovered may in fact be identical to plaquette loop current order in the OsO$_6$ cage.
We find that the magnetic field has a pattern which, appropriately, might be expected from a set of eight alternating ``magnetic monopoles'' arranged on a cube,
as shown in Fig.~4 (right panel), to form an octupole centered on the Os$^{6+}$ ion.
Fig.~\ref{fig:magfield} shows the magnetic field expected from these orbital currents
as a function of distance from the Os$^{6+}$ ion along the $[111]$ direction, where the field strength is the largest, 
for two different choices of $r_0$ as indicated. Fig.~\ref{fig:magratio} shows the same calculation, but normalizing the field by that generated by 
a $1\mu_B$ dipole located at the Os$^{6+}$ site.

\begin{figure}[t]
\includegraphics[clip,width=0.8\columnwidth]{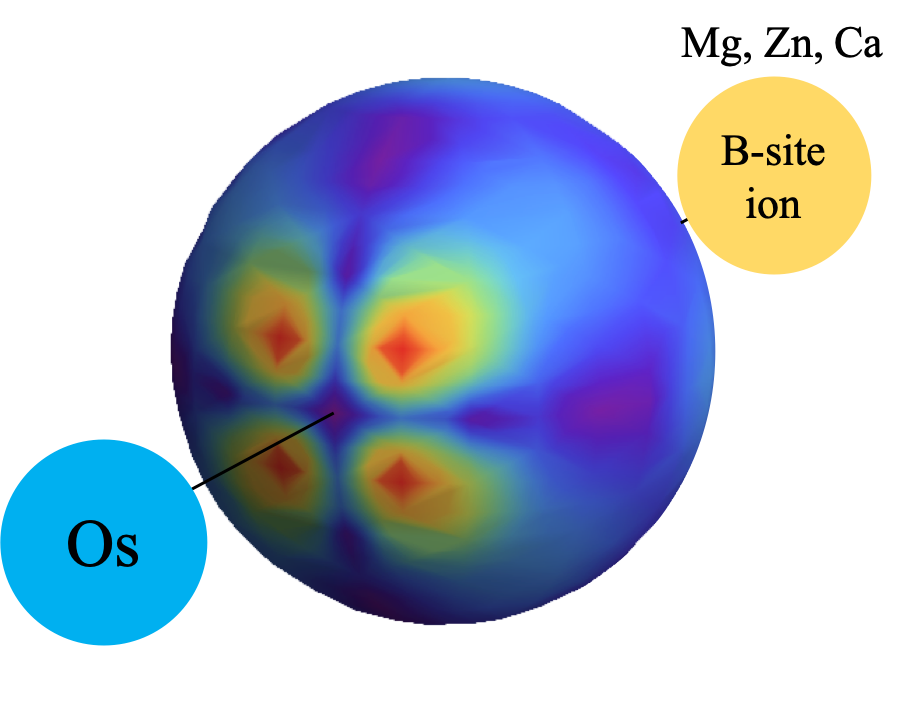}
\caption{Color plot of the ferro-octupolar magnetic field distribution over a sphere of radius $1$\AA\, around the oxygen site where the muon is expected to be bound. 
The oxygen site is located half-way between Os and the B-site ion (Mg, Zn, Ca). The largest field strength (in red) appears near the Os$^{6+}$ ion.}
\label{fig:muonfield}
\end{figure}

While we have discussed above the magnetic field due to octupolar order
as a function of distance from Os, in order to make a comparison with $\mu$SR experiments, 
we have to  estimate the fields produced by the octupolar order at possible muon stopping sites. We thus next estimate the magnetic field distribution over 
the surface of a sphere of radius $1$\AA\, centered around the oxygen site, which is where the muon is expected to be bound \cite{Dawson1988,Blundell2015}. 
Fig.~\ref{fig:muonfield} shows a plot of the field
distribution, where we find the maximum field to be present at points on this sphere located near the Os$^{6+}$ ion. (This calculation retained $9$
Os$^{6+}$ ions closest to the oxygen ion, beyond which the contribution was negligible.) We note that these maxima lie between the 
$\la 111\ra$ and $\la 100 \ra$ directions. The presence of four symmetric maxima of the field strength is consistent with the residual
symmetry in the ferro-octupolar state of $C_4$ rotations about the Os-O axis followed by time-reversal.
The computed maximum field is found to be $\sim \! 30$ Gauss, within a factor-of-two of the $\sim\! 50$ Gauss magnetic field inferred from $\mu$SR experiments 
on Ba$_2$ZnOsO$_6$, Ba$_2$CaOsO$_6$, and Ba$_2$MgOsO$_6$ below a transition temperature $T^*$. A quantitative computation with the
$\mu$SR results would need to retain the Os-O hybridization and {\it ab initio} calculations for the optimal muon stopping sites. \cite{Dawson1988,Blundell2015}
The magnetic field inferred from $\mu$SR experiments
was previously attributed to possible weak magnetic dipolar order, with a tiny ordered moment $\lesssim 0.02\mu_B$. Such a tiny ordered moment 
is difficult to explain given the typical $\sim \! 1\mu_B$ local moments expected in such Mott insulators, unless one is fine-tuned to be near a quantum critical point. 
Our work instead naturally rules out dipolar order, and instead explains this weak field as arising from loop currents in a phase which supports octupolar order.

%\begin{figure}[t]\includegraphics[width=0.49\textwidth]{Muonfield.png}
%  \caption{Schematic color plot of the ferro-octupolar magnetic field distribution over a sphere of radius $1$\AA\, around the oxygen site, located half-way between 
%  Os and the B-site ion (Mg, Zn, Ca), where the muon is expected to be bound. The largest field strength (in red) appears near the Os$^{6+}$ ion. 
%  Our calculation shows that this maximal value is $\sim\! 30$ Gauss, comparable to the measured value $\sim\!60$ Gauss from $\mu$SR experiments.}
%  \label{fig:muonfield}
%\end{figure}

\section{Reversed level scheme: Magnetic triplet ground state}

In previous work and in the above sections, we have extensively explored the case where the $J\!=\!2$ multiplet is split into a low-energy
non-Kramers doublet and a spin-gapped magnetic triplet. In this section, we explore the single-ion physics of the reversed level scheme
which has also not been studied in the oxides literature. As an illustrative example of a model which leads to this level
ordering, we explore the Hamiltonian in Eq.~\ref{eq:hfull}, but with $\delta U' \!=\! U' \!-\! (U \!-\! 2 J_H) \! < \!0$, and projecting onto
just the $t_{2g}$ orbitals. We note that this deviation is not necessarily
the only way in which the Coulomb interaction can deviate from spherical symmetry --- indeed, imposing only the octahedral point group 
symmetry will allow for a broader set of interactions.

\begin{figure}[t]
  \includegraphics[width=0.5\textwidth]{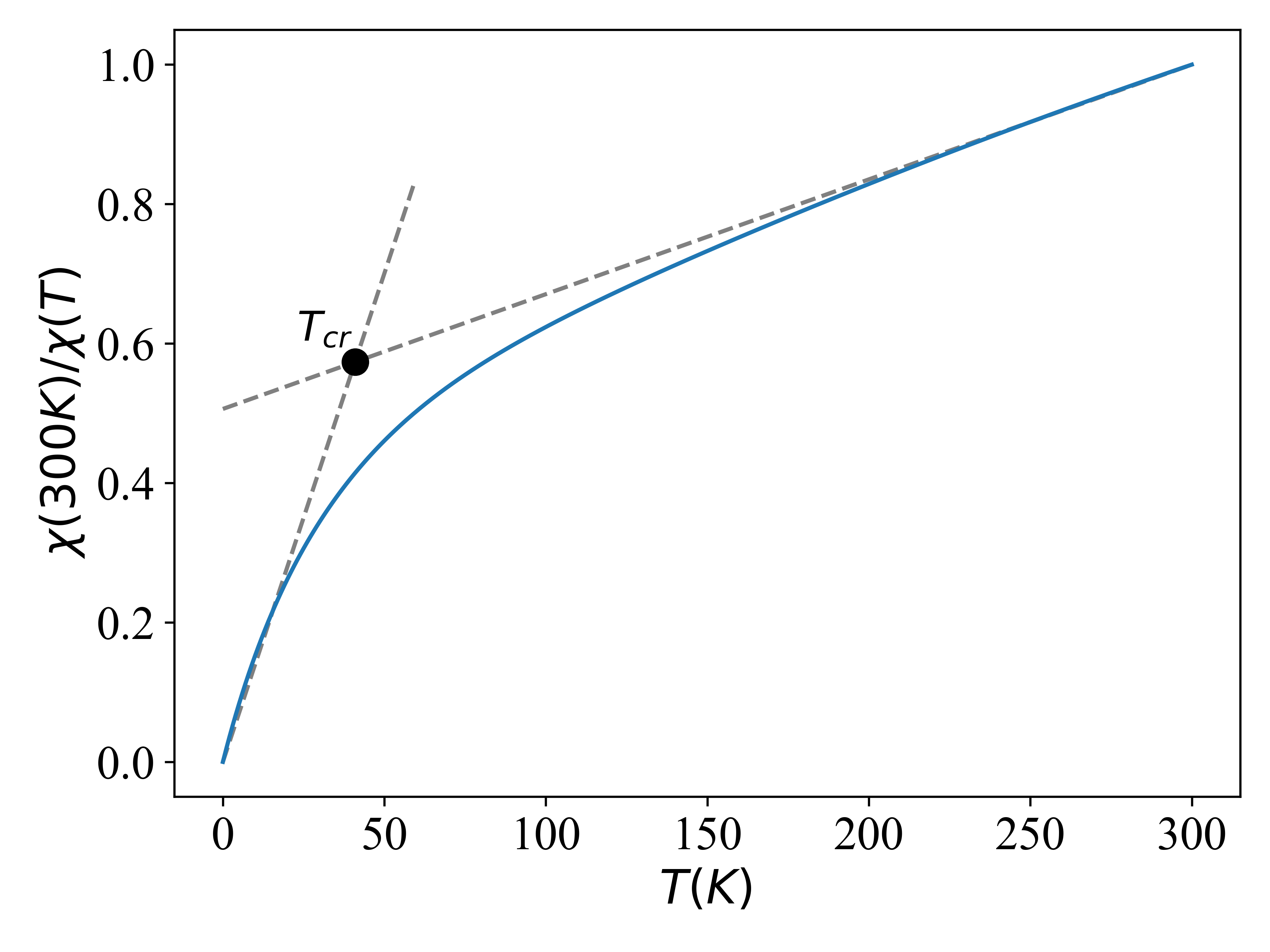}
  \caption{Temperature dependence of the inverse magnetic susceptibility (normalized to its value at $T\!=\!300$\,K) 
  in the single-site problem with a low lying magnetic triplet and an excited non-Kramers doublet; see text for details
  of the model and parameters.
  At high temperature, we find a ``Curie-Weiss''-like linear form
  $\chi^{-1}(T) \propto (T + T_s)$, as indicated by the dashed line, with $T_s \! \sim  275$\,K for the chosen parameters. 
  At low temperature, we find the Curie law  $\chi^{-1}(T) \propto T$.
  The temperature where the low $T$ and high $T$ lines meet denotes a crossover temperature scale $T_{cr} \!\approx\! 30$\,K. 
Varying the doublet-triplet splitting, we find that $k_B T_{cr} \!\approx \!0.07 |\Delta|$ and 
  $k_B T_s \!\approx \! 0.35|\Delta|$.}
  \label{fig:Re}
\end{figure}

Fig.~\ref{fig:Re} shows the inverse magnetic susceptibility $\chi^{-1}(T)$ in this single-ion case,
normalized by its value at $T\!=\! 300$\,K, for a choice
of parameters $V_C \! = \! 3$\,eV, $U \! = \! 2.5$\,eV, $\lambda \! = \! 0.4$\,eV, 
and $J_H \! = \! 0.25$\,eV (as used in the previous sections), but with $\delta U' \!=\! - 0.5$\,eV. 
(This choice of an admittedly large $\delta U'$ is only used for the simplest model to illustrate the impact of splitting 
the lowest energy $J\!=\!2$ multiplet; it is not meant to capture the full spectrum of higher energy excitations.)
This leads to a triplet ground state, with
an excited non-Kramers doublet at an energy $|\Delta| \sim\! 37$\,meV.
Interestingly, we find that $\chi^{-1}(T) \propto (T + T_{\rm s})$ in this case, exhibiting
an apparent ``Curie-Weiss''-like form with $T_{\rm s} \approx 275$\,K, over a wide range of temperatures $\gtrsim \! 150$\,K. Based on this, one might misleadingly 
infer a Curie-Weiss temperature $ \sim\! - \! 275$\,K. Only upon going to lower temperatures, do we observe a change of slope 
and the correct $\chi^{-1}(T) \propto T$ Curie law associated with the
single-ion low energy magnetic triplet. We find a very similar result in an even simpler model where we split the $J\!=\!2$ multiplet using
symmetry-allowed Stevens operators, via $H_{\rm eff} = - V_{\rm eff} ({\cal O}_{40} + 5  {\cal O}_{44})$, with $V_{\rm eff} < 0$, where
\bea
{\cal O}_{40} &=& 35 J^4_z - (30 J (J+1)-25) J_z^2 + 3 J^2(J+1)^2 \nonumber \\
&-& 6 J (J+1), \\
{\cal O}_{44} &=& \frac{1}{2} (J_+^4  + J_-^4),
\eea
suggesting that it is a robust consequence of triplet-doublet splitting, with $T_{\rm s}$ reflecting single-ion physics; in this model, $|\Delta| = 120 |V_{\rm eff}|$.
Varying the doublet-triplet splitting,
we find $k_B T_s \!\approx\! 0.35 |\Delta|$,
while the crossover from the high temperature ``Curie-Weiss-like'' form to the low temperature behavior occurs at a temperature $T_{cr}$ given
by $k_B T_{cr} \!\approx\! 0.07 |\Delta|$.

Remarkably, precisely such a behavior, with a Curie-Weiss-like form for $\chi^{-1}(T)$ and a break in slope on going below $\lesssim 150$\,K has been observed
\cite{Aharen_PRB2010} in Ba$_2$YReO$_6$,
leading us to suspect that the experimentally reported large ``Curie-Weiss'' temperature $\sim -600$\,K may in fact be misleading, and
could partly reflect this modified single-ion physics. The true Curie-Weiss temperature in this material may thus well be much smaller, and likely closer to that seen in the 
$d^2$ osmates discussed above. Our exploration thus serves to partly rationalize the widely diverging ``Curie-Weiss'' temperatures reported in this class of 
materials as arising from the differences in the single-ion physics of different $5d$ ions.
The nature and strength of exchange interactions between such magnetic ions will be discussed
elsewhere, in the context of ongoing experiments on Ba$_2$YReO$_6$.

\section{Discussion}
We have shown that the physics of spin-orbit coupled $J\!=\!2$ magnets can exhibit unconventional multipolar orders which emerge
from a low energy non-Kramers doublet. This doublet arises from crystal field splitting of the $J=2$ multiplet due to
multiple physical effects: weak $t_{2g}$-$e_g$ mixing as well as deviation of the Coulomb interaction from spherical symmetry.
Ferro-octupolar ordering within this doublet, which can result from the interplay of magnetic exchange and orbital repulsion 
\cite{paramekanti2019octupolar}, provides the most viable explanation for the huge body of experimental data, including the
$\mu$SR oscillations which we have shown results from orbital electronic currents.
As a further test of our theory, we propose that nuclear magnetic resonance (NMR) studies on the oxygen site should show no
sign of any internal fields below $T^*$ due to its octupolar structure, which is evident from the schematic plot in Fig.~4; specifically,
the octupolar configuration in a cubic system is invariant under $C_4$ rotations about the Os-O axis followed by time-reversal. This vanishing
of the field in oxygen NMR would serve to further distinguish octupolar order from possible dipolar order for which we do expect to 
see an internal field in the NMR spectrum. Applying uniaxial pressure along the $\la 111 \ra$ or $\la 110\ra$ directions would break this 
$C_4$ symmetry, leading to a nonzero field at the oxygen site which may be detectable by NMR. In previous work,\cite{paramekanti2019octupolar} 
we have also shown
how Raman scattering in a $\la 111 \ra$ magnetic field can uncover octupolar order via the appearance of new modes below $T^*$.
Our work makes a compelling case for octupolar order in a $d$-orbital Mott insulator.
Future experimental studies using pressure or doping,
to suppress the octupolar transition temperature and induce metallicity, may allow one to study possible non-Fermi liquid states 
associated with  fluctuating multipolar orders \cite{Patri2020NFL}.
Our work emphasizes the need for additional {\it ab initio} studies of $5d$ oxides at various filling factors 
to construct the appropriate Wannier functions in order to extract the local interaction Hamiltonian.
In light of our work, it is also imperative to revisit the
entire body of experiments on other $5d^2$ materials, such as Ba$_2$YReO$_6$, as well as $5d$ oxides at other filling factors.

\section{Acknowledgments}
This work was supported by the Natural Sciences and Engineering Research Council of Canada.

\bibliography{octupolar}

\begin{thebibliography}{37}%
\makeatletter
\providecommand \@ifxundefined [1]{%
 \@ifx{#1\undefined}
}%
\providecommand \@ifnum [1]{%
 \ifnum #1\expandafter \@firstoftwo
 \else \expandafter \@secondoftwo
 \fi
}%
\providecommand \@ifx [1]{%
 \ifx #1\expandafter \@firstoftwo
 \else \expandafter \@secondoftwo
 \fi
}%
\providecommand \natexlab [1]{#1}%
\providecommand \enquote  [1]{``#1''}%
\providecommand \bibnamefont  [1]{#1}%
\providecommand \bibfnamefont [1]{#1}%
\providecommand \citenamefont [1]{#1}%
\providecommand \href@noop [0]{\@secondoftwo}%
\providecommand \href [0]{\begingroup \@sanitize@url \@href}%
\providecommand \@href[1]{\@@startlink{#1}\@@href}%
\providecommand \@@href[1]{\endgroup#1\@@endlink}%
\providecommand \@sanitize@url [0]{\catcode `\\12\catcode `\$12\catcode
  `\&12\catcode `\#12\catcode `\^12\catcode `\_12\catcode `\%12\relax}%
\providecommand \@@startlink[1]{}%
\providecommand \@@endlink[0]{}%
\providecommand \url  [0]{\begingroup\@sanitize@url \@url }%
\providecommand \@url [1]{\endgroup\@href {#1}{\urlprefix }}%
\providecommand \urlprefix  [0]{URL }%
\providecommand \Eprint [0]{\href }%
\providecommand \doibase [0]{http://dx.doi.org/}%
\providecommand \selectlanguage [0]{\@gobble}%
\providecommand \bibinfo  [0]{\@secondoftwo}%
\providecommand \bibfield  [0]{\@secondoftwo}%
\providecommand \translation [1]{[#1]}%
\providecommand \BibitemOpen [0]{}%
\providecommand \bibitemStop [0]{}%
\providecommand \bibitemNoStop [0]{.\EOS\space}%
\providecommand \EOS [0]{\spacefactor3000\relax}%
\providecommand \BibitemShut  [1]{\csname bibitem#1\endcsname}%
\let\auto@bib@innerbib\@empty
%</preamble>
\bibitem [{\citenamefont {Santini}\ \emph {et~al.}(2009)\citenamefont
  {Santini}, \citenamefont {Carretta}, \citenamefont {Amoretti}, \citenamefont
  {Caciuffo}, \citenamefont {Magnani},\ and\ \citenamefont
  {Lander}}]{MultipolarRMP2009}%
  \BibitemOpen
  \bibfield  {author} {\bibinfo {author} {\bibfnamefont {P.}~\bibnamefont
  {Santini}}, \bibinfo {author} {\bibfnamefont {S.}~\bibnamefont {Carretta}},
  \bibinfo {author} {\bibfnamefont {G.}~\bibnamefont {Amoretti}}, \bibinfo
  {author} {\bibfnamefont {R.}~\bibnamefont {Caciuffo}}, \bibinfo {author}
  {\bibfnamefont {N.}~\bibnamefont {Magnani}}, \ and\ \bibinfo {author}
  {\bibfnamefont {G.~H.}\ \bibnamefont {Lander}},\ }\href {\doibase
  10.1103/RevModPhys.81.807} {\bibfield  {journal} {\bibinfo  {journal} {Rev.
  Mod. Phys.}\ }\textbf {\bibinfo {volume} {81}},\ \bibinfo {pages} {807}
  (\bibinfo {year} {2009})}\BibitemShut {NoStop}%
\bibitem [{\citenamefont {Haule}\ and\ \citenamefont
  {Kotliar}(2009)}]{HauleKotliar2009}%
  \BibitemOpen
  \bibfield  {author} {\bibinfo {author} {\bibfnamefont {K.}~\bibnamefont
  {Haule}}\ and\ \bibinfo {author} {\bibfnamefont {G.}~\bibnamefont
  {Kotliar}},\ }\href {https://doi.org/10.1038/nphys1392} {\bibfield  {journal}
  {\bibinfo  {journal} {Nature Physics}\ }\textbf {\bibinfo {volume} {5}},\
  \bibinfo {pages} {796 EP } (\bibinfo {year} {2009})}\BibitemShut {NoStop}%
\bibitem [{\citenamefont {Santini}\ and\ \citenamefont
  {Amoretti}(2000)}]{SantiniNpO2_PRL2000}%
  \BibitemOpen
  \bibfield  {author} {\bibinfo {author} {\bibfnamefont {P.}~\bibnamefont
  {Santini}}\ and\ \bibinfo {author} {\bibfnamefont {G.}~\bibnamefont
  {Amoretti}},\ }\href {\doibase 10.1103/PhysRevLett.85.2188} {\bibfield
  {journal} {\bibinfo  {journal} {Phys. Rev. Lett.}\ }\textbf {\bibinfo
  {volume} {85}},\ \bibinfo {pages} {2188} (\bibinfo {year}
  {2000})}\BibitemShut {NoStop}%
\bibitem [{\citenamefont {Paix\~ao}\ \emph {et~al.}(2002)\citenamefont
  {Paix\~ao}, \citenamefont {Detlefs}, \citenamefont {Longfield}, \citenamefont
  {Caciuffo}, \citenamefont {Santini}, \citenamefont {Bernhoeft}, \citenamefont
  {Rebizant},\ and\ \citenamefont {Lander}}]{NpO2TripleQ_PRL2002}%
  \BibitemOpen
  \bibfield  {author} {\bibinfo {author} {\bibfnamefont {J.~A.}\ \bibnamefont
  {Paix\~ao}}, \bibinfo {author} {\bibfnamefont {C.}~\bibnamefont {Detlefs}},
  \bibinfo {author} {\bibfnamefont {M.~J.}\ \bibnamefont {Longfield}}, \bibinfo
  {author} {\bibfnamefont {R.}~\bibnamefont {Caciuffo}}, \bibinfo {author}
  {\bibfnamefont {P.}~\bibnamefont {Santini}}, \bibinfo {author} {\bibfnamefont
  {N.}~\bibnamefont {Bernhoeft}}, \bibinfo {author} {\bibfnamefont
  {J.}~\bibnamefont {Rebizant}}, \ and\ \bibinfo {author} {\bibfnamefont
  {G.~H.}\ \bibnamefont {Lander}},\ }\href {\doibase
  10.1103/PhysRevLett.89.187202} {\bibfield  {journal} {\bibinfo  {journal}
  {Phys. Rev. Lett.}\ }\textbf {\bibinfo {volume} {89}},\ \bibinfo {pages}
  {187202} (\bibinfo {year} {2002})}\BibitemShut {NoStop}%
\bibitem [{\citenamefont {Kiss}\ and\ \citenamefont
  {Fazekas}(2003)}]{Fazekas_PRB2003}%
  \BibitemOpen
  \bibfield  {author} {\bibinfo {author} {\bibfnamefont {A.}~\bibnamefont
  {Kiss}}\ and\ \bibinfo {author} {\bibfnamefont {P.}~\bibnamefont {Fazekas}},\
  }\href {\doibase 10.1103/PhysRevB.68.174425} {\bibfield  {journal} {\bibinfo
  {journal} {Phys. Rev. B}\ }\textbf {\bibinfo {volume} {68}},\ \bibinfo
  {pages} {174425} (\bibinfo {year} {2003})}\BibitemShut {NoStop}%
\bibitem [{\citenamefont {Tokunaga}\ \emph {et~al.}(2006)\citenamefont
  {Tokunaga}, \citenamefont {Aoki}, \citenamefont {Homma}, \citenamefont
  {Kambe}, \citenamefont {Sakai}, \citenamefont {Ikeda}, \citenamefont
  {Fujimoto}, \citenamefont {Walstedt}, \citenamefont {Yasuoka}, \citenamefont
  {Yamamoto}, \citenamefont {Nakamura},\ and\ \citenamefont
  {Shiokawa}}]{NpO2NMR_PRL2006}%
  \BibitemOpen
  \bibfield  {author} {\bibinfo {author} {\bibfnamefont {Y.}~\bibnamefont
  {Tokunaga}}, \bibinfo {author} {\bibfnamefont {D.}~\bibnamefont {Aoki}},
  \bibinfo {author} {\bibfnamefont {Y.}~\bibnamefont {Homma}}, \bibinfo
  {author} {\bibfnamefont {S.}~\bibnamefont {Kambe}}, \bibinfo {author}
  {\bibfnamefont {H.}~\bibnamefont {Sakai}}, \bibinfo {author} {\bibfnamefont
  {S.}~\bibnamefont {Ikeda}}, \bibinfo {author} {\bibfnamefont
  {T.}~\bibnamefont {Fujimoto}}, \bibinfo {author} {\bibfnamefont {R.~E.}\
  \bibnamefont {Walstedt}}, \bibinfo {author} {\bibfnamefont {H.}~\bibnamefont
  {Yasuoka}}, \bibinfo {author} {\bibfnamefont {E.}~\bibnamefont {Yamamoto}},
  \bibinfo {author} {\bibfnamefont {A.}~\bibnamefont {Nakamura}}, \ and\
  \bibinfo {author} {\bibfnamefont {Y.}~\bibnamefont {Shiokawa}},\ }\href
  {\doibase 10.1103/PhysRevLett.97.257601} {\bibfield  {journal} {\bibinfo
  {journal} {Phys. Rev. Lett.}\ }\textbf {\bibinfo {volume} {97}},\ \bibinfo
  {pages} {257601} (\bibinfo {year} {2006})}\BibitemShut {NoStop}%
\bibitem [{\citenamefont {Arima}(2013)}]{Arima2013}%
  \BibitemOpen
  \bibfield  {author} {\bibinfo {author} {\bibfnamefont {T.-h.}\ \bibnamefont
  {Arima}},\ }\href {\doibase 10.7566/JPSJ.82.013705} {\bibfield  {journal}
  {\bibinfo  {journal} {Journal of the Physical Society of Japan}\ }\textbf
  {\bibinfo {volume} {82}},\ \bibinfo {pages} {013705} (\bibinfo {year}
  {2013})}\BibitemShut {NoStop}%
\bibitem [{\citenamefont {Sakai}\ and\ \citenamefont
  {Nakatsuji}(2011)}]{Sakai_JPSJ2011}%
  \BibitemOpen
  \bibfield  {author} {\bibinfo {author} {\bibfnamefont {A.}~\bibnamefont
  {Sakai}}\ and\ \bibinfo {author} {\bibfnamefont {S.}~\bibnamefont
  {Nakatsuji}},\ }\href {\doibase 10.1143/JPSJ.80.063701} {\bibfield  {journal}
  {\bibinfo  {journal} {Journal of the Physical Society of Japan}\ }\textbf
  {\bibinfo {volume} {80}},\ \bibinfo {pages} {063701} (\bibinfo {year}
  {2011})}\BibitemShut {NoStop}%
\bibitem [{\citenamefont {Sato}\ \emph {et~al.}(2012)\citenamefont {Sato},
  \citenamefont {Ibuka}, \citenamefont {Nambu}, \citenamefont {Yamazaki},
  \citenamefont {Hong}, \citenamefont {Sakai},\ and\ \citenamefont
  {Nakatsuji}}]{Sato_PRB2012}%
  \BibitemOpen
  \bibfield  {author} {\bibinfo {author} {\bibfnamefont {T.~J.}\ \bibnamefont
  {Sato}}, \bibinfo {author} {\bibfnamefont {S.}~\bibnamefont {Ibuka}},
  \bibinfo {author} {\bibfnamefont {Y.}~\bibnamefont {Nambu}}, \bibinfo
  {author} {\bibfnamefont {T.}~\bibnamefont {Yamazaki}}, \bibinfo {author}
  {\bibfnamefont {T.}~\bibnamefont {Hong}}, \bibinfo {author} {\bibfnamefont
  {A.}~\bibnamefont {Sakai}}, \ and\ \bibinfo {author} {\bibfnamefont
  {S.}~\bibnamefont {Nakatsuji}},\ }\href {\doibase 10.1103/PhysRevB.86.184419}
  {\bibfield  {journal} {\bibinfo  {journal} {Phys. Rev. B}\ }\textbf {\bibinfo
  {volume} {86}},\ \bibinfo {pages} {184419} (\bibinfo {year}
  {2012})}\BibitemShut {NoStop}%
\bibitem [{\citenamefont {Tsujimoto}\ \emph {et~al.}(2014)\citenamefont
  {Tsujimoto}, \citenamefont {Matsumoto}, \citenamefont {Tomita}, \citenamefont
  {Sakai},\ and\ \citenamefont {Nakatsuji}}]{Nakatsuji_PRL2014}%
  \BibitemOpen
  \bibfield  {author} {\bibinfo {author} {\bibfnamefont {M.}~\bibnamefont
  {Tsujimoto}}, \bibinfo {author} {\bibfnamefont {Y.}~\bibnamefont
  {Matsumoto}}, \bibinfo {author} {\bibfnamefont {T.}~\bibnamefont {Tomita}},
  \bibinfo {author} {\bibfnamefont {A.}~\bibnamefont {Sakai}}, \ and\ \bibinfo
  {author} {\bibfnamefont {S.}~\bibnamefont {Nakatsuji}},\ }\href {\doibase
  10.1103/PhysRevLett.113.267001} {\bibfield  {journal} {\bibinfo  {journal}
  {Phys. Rev. Lett.}\ }\textbf {\bibinfo {volume} {113}},\ \bibinfo {pages}
  {267001} (\bibinfo {year} {2014})}\BibitemShut {NoStop}%
\bibitem [{\citenamefont {Hattori}\ and\ \citenamefont
  {Tsunetsugu}(2016)}]{Hattori2016}%
  \BibitemOpen
  \bibfield  {author} {\bibinfo {author} {\bibfnamefont {K.}~\bibnamefont
  {Hattori}}\ and\ \bibinfo {author} {\bibfnamefont {H.}~\bibnamefont
  {Tsunetsugu}},\ }\href {\doibase 10.7566/JPSJ.85.094001} {\bibfield
  {journal} {\bibinfo  {journal} {Journal of the Physical Society of Japan}\
  }\textbf {\bibinfo {volume} {85}},\ \bibinfo {pages} {094001} (\bibinfo
  {year} {2016})}\BibitemShut {NoStop}%
\bibitem [{\citenamefont {Freyer}\ \emph {et~al.}(2018)\citenamefont {Freyer},
  \citenamefont {Attig}, \citenamefont {Lee}, \citenamefont {Paramekanti},
  \citenamefont {Trebst},\ and\ \citenamefont {Kim}}]{Freyer2018}%
  \BibitemOpen
  \bibfield  {author} {\bibinfo {author} {\bibfnamefont {F.}~\bibnamefont
  {Freyer}}, \bibinfo {author} {\bibfnamefont {J.}~\bibnamefont {Attig}},
  \bibinfo {author} {\bibfnamefont {S.}~\bibnamefont {Lee}}, \bibinfo {author}
  {\bibfnamefont {A.}~\bibnamefont {Paramekanti}}, \bibinfo {author}
  {\bibfnamefont {S.}~\bibnamefont {Trebst}}, \ and\ \bibinfo {author}
  {\bibfnamefont {Y.~B.}\ \bibnamefont {Kim}},\ }\href {\doibase
  10.1103/PhysRevB.97.115111} {\bibfield  {journal} {\bibinfo  {journal} {Phys.
  Rev. B}\ }\textbf {\bibinfo {volume} {97}},\ \bibinfo {pages} {115111}
  (\bibinfo {year} {2018})}\BibitemShut {NoStop}%
\bibitem [{\citenamefont {Lee}\ \emph {et~al.}(2018)\citenamefont {Lee},
  \citenamefont {Trebst}, \citenamefont {Kim},\ and\ \citenamefont
  {Paramekanti}}]{SBLee2018}%
  \BibitemOpen
  \bibfield  {author} {\bibinfo {author} {\bibfnamefont {S.}~\bibnamefont
  {Lee}}, \bibinfo {author} {\bibfnamefont {S.}~\bibnamefont {Trebst}},
  \bibinfo {author} {\bibfnamefont {Y.~B.}\ \bibnamefont {Kim}}, \ and\
  \bibinfo {author} {\bibfnamefont {A.}~\bibnamefont {Paramekanti}},\ }\href
  {\doibase 10.1103/PhysRevB.98.134447} {\bibfield  {journal} {\bibinfo
  {journal} {Phys. Rev. B}\ }\textbf {\bibinfo {volume} {98}},\ \bibinfo
  {pages} {134447} (\bibinfo {year} {2018})}\BibitemShut {NoStop}%
\bibitem [{\citenamefont {Patri}\ \emph {et~al.}(2019)\citenamefont {Patri},
  \citenamefont {Sakai}, \citenamefont {Lee}, \citenamefont {Paramekanti},
  \citenamefont {Nakatsuji},\ and\ \citenamefont {Kim}}]{Patri2019}%
  \BibitemOpen
  \bibfield  {author} {\bibinfo {author} {\bibfnamefont {A.~S.}\ \bibnamefont
  {Patri}}, \bibinfo {author} {\bibfnamefont {A.}~\bibnamefont {Sakai}},
  \bibinfo {author} {\bibfnamefont {S.}~\bibnamefont {Lee}}, \bibinfo {author}
  {\bibfnamefont {A.}~\bibnamefont {Paramekanti}}, \bibinfo {author}
  {\bibfnamefont {S.}~\bibnamefont {Nakatsuji}}, \ and\ \bibinfo {author}
  {\bibfnamefont {Y.~B.}\ \bibnamefont {Kim}},\ }\href {\doibase
  10.1038/s41467-019-11913-3} {\bibfield  {journal} {\bibinfo  {journal}
  {Nature Communications}\ }\textbf {\bibinfo {volume} {10}},\ \bibinfo {pages}
  {4092} (\bibinfo {year} {2019})}\BibitemShut {NoStop}%
\bibitem [{\citenamefont {Fu}(2015)}]{LFu_PRL2015}%
  \BibitemOpen
  \bibfield  {author} {\bibinfo {author} {\bibfnamefont {L.}~\bibnamefont
  {Fu}},\ }\href {\doibase 10.1103/PhysRevLett.115.026401} {\bibfield
  {journal} {\bibinfo  {journal} {Phys. Rev. Lett.}\ }\textbf {\bibinfo
  {volume} {115}},\ \bibinfo {pages} {026401} (\bibinfo {year}
  {2015})}\BibitemShut {NoStop}%
\bibitem [{\citenamefont {Harter}\ \emph {et~al.}(2017)\citenamefont {Harter},
  \citenamefont {Zhao}, \citenamefont {Yan}, \citenamefont {Mandrus},\ and\
  \citenamefont {Hsieh}}]{Hsieh_Science2017}%
  \BibitemOpen
  \bibfield  {author} {\bibinfo {author} {\bibfnamefont {J.~W.}\ \bibnamefont
  {Harter}}, \bibinfo {author} {\bibfnamefont {Z.~Y.}\ \bibnamefont {Zhao}},
  \bibinfo {author} {\bibfnamefont {J.-Q.}\ \bibnamefont {Yan}}, \bibinfo
  {author} {\bibfnamefont {D.~G.}\ \bibnamefont {Mandrus}}, \ and\ \bibinfo
  {author} {\bibfnamefont {D.}~\bibnamefont {Hsieh}},\ }\href {\doibase
  10.1126/science.aad1188} {\bibfield  {journal} {\bibinfo  {journal}
  {Science}\ }\textbf {\bibinfo {volume} {356}},\ \bibinfo {pages} {295}
  (\bibinfo {year} {2017})}\BibitemShut {NoStop}%
\bibitem [{\citenamefont {Hayami}\ \emph {et~al.}(2018)\citenamefont {Hayami},
  \citenamefont {Kusunose},\ and\ \citenamefont {Motome}}]{Motome2018}%
  \BibitemOpen
  \bibfield  {author} {\bibinfo {author} {\bibfnamefont {S.}~\bibnamefont
  {Hayami}}, \bibinfo {author} {\bibfnamefont {H.}~\bibnamefont {Kusunose}}, \
  and\ \bibinfo {author} {\bibfnamefont {Y.}~\bibnamefont {Motome}},\ }\href
  {\doibase 10.1103/PhysRevB.97.024414} {\bibfield  {journal} {\bibinfo
  {journal} {Phys. Rev. B}\ }\textbf {\bibinfo {volume} {97}},\ \bibinfo
  {pages} {024414} (\bibinfo {year} {2018})}\BibitemShut {NoStop}%
\bibitem [{\citenamefont {Maharaj}\ \emph {et~al.}(2020)\citenamefont
  {Maharaj}, \citenamefont {Sala}, \citenamefont {Stone}, \citenamefont
  {Kermarrec}, \citenamefont {Ritter}, \citenamefont {Fauth}, \citenamefont
  {Marjerrison}, \citenamefont {Greedan}, \citenamefont {Paramekanti},\ and\
  \citenamefont {Gaulin}}]{maharaj2019octupolar}%
  \BibitemOpen
  \bibfield  {author} {\bibinfo {author} {\bibfnamefont {D.~D.}\ \bibnamefont
  {Maharaj}}, \bibinfo {author} {\bibfnamefont {G.}~\bibnamefont {Sala}},
  \bibinfo {author} {\bibfnamefont {M.~B.}\ \bibnamefont {Stone}}, \bibinfo
  {author} {\bibfnamefont {E.}~\bibnamefont {Kermarrec}}, \bibinfo {author}
  {\bibfnamefont {C.}~\bibnamefont {Ritter}}, \bibinfo {author} {\bibfnamefont
  {F.}~\bibnamefont {Fauth}}, \bibinfo {author} {\bibfnamefont {C.~A.}\
  \bibnamefont {Marjerrison}}, \bibinfo {author} {\bibfnamefont {J.~E.}\
  \bibnamefont {Greedan}}, \bibinfo {author} {\bibfnamefont {A.}~\bibnamefont
  {Paramekanti}}, \ and\ \bibinfo {author} {\bibfnamefont {B.~D.}\ \bibnamefont
  {Gaulin}},\ }\href {\doibase 10.1103/PhysRevLett.124.087206} {\bibfield
  {journal} {\bibinfo  {journal} {Phys. Rev. Lett.}\ }\textbf {\bibinfo
  {volume} {124}},\ \bibinfo {pages} {087206} (\bibinfo {year}
  {2020})}\BibitemShut {NoStop}%
\bibitem [{\citenamefont {Paramekanti}\ \emph {et~al.}(2020)\citenamefont
  {Paramekanti}, \citenamefont {Maharaj},\ and\ \citenamefont
  {Gaulin}}]{paramekanti2019octupolar}%
  \BibitemOpen
  \bibfield  {author} {\bibinfo {author} {\bibfnamefont {A.}~\bibnamefont
  {Paramekanti}}, \bibinfo {author} {\bibfnamefont {D.~D.}\ \bibnamefont
  {Maharaj}}, \ and\ \bibinfo {author} {\bibfnamefont {B.~D.}\ \bibnamefont
  {Gaulin}},\ }\href {\doibase 10.1103/PhysRevB.101.054439} {\bibfield
  {journal} {\bibinfo  {journal} {Phys. Rev. B}\ }\textbf {\bibinfo {volume}
  {101}},\ \bibinfo {pages} {054439} (\bibinfo {year} {2020})}\BibitemShut
  {NoStop}%
\bibitem [{\citenamefont {Chen}\ \emph {et~al.}(2010)\citenamefont {Chen},
  \citenamefont {Pereira},\ and\ \citenamefont {Balents}}]{ChenBalents2010}%
  \BibitemOpen
  \bibfield  {author} {\bibinfo {author} {\bibfnamefont {G.}~\bibnamefont
  {Chen}}, \bibinfo {author} {\bibfnamefont {R.}~\bibnamefont {Pereira}}, \
  and\ \bibinfo {author} {\bibfnamefont {L.}~\bibnamefont {Balents}},\ }\href
  {\doibase 10.1103/PhysRevB.82.174440} {\bibfield  {journal} {\bibinfo
  {journal} {Phys. Rev. B}\ }\textbf {\bibinfo {volume} {82}},\ \bibinfo
  {pages} {174440} (\bibinfo {year} {2010})}\BibitemShut {NoStop}%
\bibitem [{\citenamefont {Chen}\ and\ \citenamefont
  {Balents}(2011)}]{ChenBalents2011}%
  \BibitemOpen
  \bibfield  {author} {\bibinfo {author} {\bibfnamefont {G.}~\bibnamefont
  {Chen}}\ and\ \bibinfo {author} {\bibfnamefont {L.}~\bibnamefont {Balents}},\
  }\href {\doibase 10.1103/PhysRevB.84.094420} {\bibfield  {journal} {\bibinfo
  {journal} {Phys. Rev. B}\ }\textbf {\bibinfo {volume} {84}},\ \bibinfo
  {pages} {094420} (\bibinfo {year} {2011})}\BibitemShut {NoStop}%
\bibitem [{\citenamefont {Svoboda}\ \emph {et~al.}(2017)\citenamefont
  {Svoboda}, \citenamefont {Randeria},\ and\ \citenamefont
  {Trivedi}}]{Svoboda_PRB2017}%
  \BibitemOpen
  \bibfield  {author} {\bibinfo {author} {\bibfnamefont {C.}~\bibnamefont
  {Svoboda}}, \bibinfo {author} {\bibfnamefont {M.}~\bibnamefont {Randeria}}, \
  and\ \bibinfo {author} {\bibfnamefont {N.}~\bibnamefont {Trivedi}},\ }\href
  {\doibase 10.1103/PhysRevB.95.014409} {\bibfield  {journal} {\bibinfo
  {journal} {Phys. Rev. B}\ }\textbf {\bibinfo {volume} {95}},\ \bibinfo
  {pages} {014409} (\bibinfo {year} {2017})}\BibitemShut {NoStop}%
\bibitem [{\citenamefont {Thompson}\ \emph {et~al.}(2014)\citenamefont
  {Thompson}, \citenamefont {Carlo}, \citenamefont {Flacau}, \citenamefont
  {Aharen}, \citenamefont {Leahy}, \citenamefont {Pollichemi}, \citenamefont
  {Munsie}, \citenamefont {Medina}, \citenamefont {Luke}, \citenamefont
  {Munevar}, \citenamefont {Cheung}, \citenamefont {Goko}, \citenamefont
  {Uemura},\ and\ \citenamefont {Greedan}}]{Thompson_JPCM2014}%
  \BibitemOpen
  \bibfield  {author} {\bibinfo {author} {\bibfnamefont {C.~M.}\ \bibnamefont
  {Thompson}}, \bibinfo {author} {\bibfnamefont {J.~P.}\ \bibnamefont {Carlo}},
  \bibinfo {author} {\bibfnamefont {R.}~\bibnamefont {Flacau}}, \bibinfo
  {author} {\bibfnamefont {T.}~\bibnamefont {Aharen}}, \bibinfo {author}
  {\bibfnamefont {I.~A.}\ \bibnamefont {Leahy}}, \bibinfo {author}
  {\bibfnamefont {J.~R.}\ \bibnamefont {Pollichemi}}, \bibinfo {author}
  {\bibfnamefont {T.~J.~S.}\ \bibnamefont {Munsie}}, \bibinfo {author}
  {\bibfnamefont {T.}~\bibnamefont {Medina}}, \bibinfo {author} {\bibfnamefont
  {G.~M.}\ \bibnamefont {Luke}}, \bibinfo {author} {\bibfnamefont
  {J.}~\bibnamefont {Munevar}}, \bibinfo {author} {\bibfnamefont
  {S.}~\bibnamefont {Cheung}}, \bibinfo {author} {\bibfnamefont
  {T.}~\bibnamefont {Goko}}, \bibinfo {author} {\bibfnamefont {Y.~J.}\
  \bibnamefont {Uemura}}, \ and\ \bibinfo {author} {\bibfnamefont {J.~E.}\
  \bibnamefont {Greedan}},\ }\href {\doibase 10.1088/0953-8984/26/30/306003}
  {\bibfield  {journal} {\bibinfo  {journal} {Journal of Physics: Condensed
  Matter}\ }\textbf {\bibinfo {volume} {26}},\ \bibinfo {pages} {306003}
  (\bibinfo {year} {2014})}\BibitemShut {NoStop}%
\bibitem [{\citenamefont {Kermarrec}\ \emph {et~al.}(2015)\citenamefont
  {Kermarrec}, \citenamefont {Marjerrison}, \citenamefont {Thompson},
  \citenamefont {Maharaj}, \citenamefont {Levin}, \citenamefont {Kroeker},
  \citenamefont {Granroth}, \citenamefont {Flacau}, \citenamefont {Yamani},
  \citenamefont {Greedan},\ and\ \citenamefont {Gaulin}}]{Kermarrec2015}%
  \BibitemOpen
  \bibfield  {author} {\bibinfo {author} {\bibfnamefont {E.}~\bibnamefont
  {Kermarrec}}, \bibinfo {author} {\bibfnamefont {C.~A.}\ \bibnamefont
  {Marjerrison}}, \bibinfo {author} {\bibfnamefont {C.~M.}\ \bibnamefont
  {Thompson}}, \bibinfo {author} {\bibfnamefont {D.~D.}\ \bibnamefont
  {Maharaj}}, \bibinfo {author} {\bibfnamefont {K.}~\bibnamefont {Levin}},
  \bibinfo {author} {\bibfnamefont {S.}~\bibnamefont {Kroeker}}, \bibinfo
  {author} {\bibfnamefont {G.~E.}\ \bibnamefont {Granroth}}, \bibinfo {author}
  {\bibfnamefont {R.}~\bibnamefont {Flacau}}, \bibinfo {author} {\bibfnamefont
  {Z.}~\bibnamefont {Yamani}}, \bibinfo {author} {\bibfnamefont {J.~E.}\
  \bibnamefont {Greedan}}, \ and\ \bibinfo {author} {\bibfnamefont {B.~D.}\
  \bibnamefont {Gaulin}},\ }\href {\doibase 10.1103/PhysRevB.91.075133}
  {\bibfield  {journal} {\bibinfo  {journal} {Phys. Rev. B}\ }\textbf {\bibinfo
  {volume} {91}},\ \bibinfo {pages} {075133} (\bibinfo {year}
  {2015})}\BibitemShut {NoStop}%
\bibitem [{\citenamefont {Thompson}\ \emph {et~al.}(2016)\citenamefont
  {Thompson}, \citenamefont {Marjerrison}, \citenamefont {Sharma},
  \citenamefont {Wiebe}, \citenamefont {Maharaj}, \citenamefont {Sala},
  \citenamefont {Flacau}, \citenamefont {Hallas}, \citenamefont {Cai},
  \citenamefont {Gaulin}, \citenamefont {Luke},\ and\ \citenamefont
  {Greedan}}]{Thompson_PRB2016}%
  \BibitemOpen
  \bibfield  {author} {\bibinfo {author} {\bibfnamefont {C.~M.}\ \bibnamefont
  {Thompson}}, \bibinfo {author} {\bibfnamefont {C.~A.}\ \bibnamefont
  {Marjerrison}}, \bibinfo {author} {\bibfnamefont {A.~Z.}\ \bibnamefont
  {Sharma}}, \bibinfo {author} {\bibfnamefont {C.~R.}\ \bibnamefont {Wiebe}},
  \bibinfo {author} {\bibfnamefont {D.~D.}\ \bibnamefont {Maharaj}}, \bibinfo
  {author} {\bibfnamefont {G.}~\bibnamefont {Sala}}, \bibinfo {author}
  {\bibfnamefont {R.}~\bibnamefont {Flacau}}, \bibinfo {author} {\bibfnamefont
  {A.~M.}\ \bibnamefont {Hallas}}, \bibinfo {author} {\bibfnamefont
  {Y.}~\bibnamefont {Cai}}, \bibinfo {author} {\bibfnamefont {B.~D.}\
  \bibnamefont {Gaulin}}, \bibinfo {author} {\bibfnamefont {G.~M.}\
  \bibnamefont {Luke}}, \ and\ \bibinfo {author} {\bibfnamefont {J.~E.}\
  \bibnamefont {Greedan}},\ }\href {\doibase 10.1103/PhysRevB.93.014431}
  {\bibfield  {journal} {\bibinfo  {journal} {Phys. Rev. B}\ }\textbf {\bibinfo
  {volume} {93}},\ \bibinfo {pages} {014431} (\bibinfo {year}
  {2016})}\BibitemShut {NoStop}%
\bibitem [{\citenamefont {Marjerrison}\ \emph {et~al.}(2016)\citenamefont
  {Marjerrison}, \citenamefont {Thompson}, \citenamefont {Sharma},
  \citenamefont {Hallas}, \citenamefont {Wilson}, \citenamefont {Munsie},
  \citenamefont {Flacau}, \citenamefont {Wiebe}, \citenamefont {Gaulin},
  \citenamefont {Luke},\ and\ \citenamefont {Greedan}}]{MarjerrisonPRB2016}%
  \BibitemOpen
  \bibfield  {author} {\bibinfo {author} {\bibfnamefont {C.~A.}\ \bibnamefont
  {Marjerrison}}, \bibinfo {author} {\bibfnamefont {C.~M.}\ \bibnamefont
  {Thompson}}, \bibinfo {author} {\bibfnamefont {A.~Z.}\ \bibnamefont
  {Sharma}}, \bibinfo {author} {\bibfnamefont {A.~M.}\ \bibnamefont {Hallas}},
  \bibinfo {author} {\bibfnamefont {M.~N.}\ \bibnamefont {Wilson}}, \bibinfo
  {author} {\bibfnamefont {T.~J.~S.}\ \bibnamefont {Munsie}}, \bibinfo {author}
  {\bibfnamefont {R.}~\bibnamefont {Flacau}}, \bibinfo {author} {\bibfnamefont
  {C.~R.}\ \bibnamefont {Wiebe}}, \bibinfo {author} {\bibfnamefont {B.~D.}\
  \bibnamefont {Gaulin}}, \bibinfo {author} {\bibfnamefont {G.~M.}\
  \bibnamefont {Luke}}, \ and\ \bibinfo {author} {\bibfnamefont {J.~E.}\
  \bibnamefont {Greedan}},\ }\href {\doibase 10.1103/PhysRevB.94.134429}
  {\bibfield  {journal} {\bibinfo  {journal} {Phys. Rev. B}\ }\textbf {\bibinfo
  {volume} {94}},\ \bibinfo {pages} {134429} (\bibinfo {year}
  {2016})}\BibitemShut {NoStop}%
\bibitem [{\citenamefont {Stamokostas}\ and\ \citenamefont
  {Fiete}(2018)}]{Fiete_eg_PRB2018}%
  \BibitemOpen
  \bibfield  {author} {\bibinfo {author} {\bibfnamefont {G.~L.}\ \bibnamefont
  {Stamokostas}}\ and\ \bibinfo {author} {\bibfnamefont {G.~A.}\ \bibnamefont
  {Fiete}},\ }\href {\doibase 10.1103/PhysRevB.97.085150} {\bibfield  {journal}
  {\bibinfo  {journal} {Phys. Rev. B}\ }\textbf {\bibinfo {volume} {97}},\
  \bibinfo {pages} {085150} (\bibinfo {year} {2018})}\BibitemShut {NoStop}%
\bibitem [{\citenamefont {Ribic}\ \emph {et~al.}(2014)\citenamefont {Ribic},
  \citenamefont {Assmann}, \citenamefont {T\'oth},\ and\ \citenamefont
  {Held}}]{Ribic_PRB2014}%
  \BibitemOpen
  \bibfield  {author} {\bibinfo {author} {\bibfnamefont {T.}~\bibnamefont
  {Ribic}}, \bibinfo {author} {\bibfnamefont {E.}~\bibnamefont {Assmann}},
  \bibinfo {author} {\bibfnamefont {A.}~\bibnamefont {T\'oth}}, \ and\ \bibinfo
  {author} {\bibfnamefont {K.}~\bibnamefont {Held}},\ }\href {\doibase
  10.1103/PhysRevB.90.165105} {\bibfield  {journal} {\bibinfo  {journal} {Phys.
  Rev. B}\ }\textbf {\bibinfo {volume} {90}},\ \bibinfo {pages} {165105}
  (\bibinfo {year} {2014})}\BibitemShut {NoStop}%
\bibitem [{\citenamefont {Yuan}\ \emph {et~al.}(2017)\citenamefont {Yuan},
  \citenamefont {Clancy}, \citenamefont {Cook}, \citenamefont {Thompson},
  \citenamefont {Greedan}, \citenamefont {Cao}, \citenamefont {Jeon},
  \citenamefont {Noh}, \citenamefont {Upton}, \citenamefont {Casa},
  \citenamefont {Gog}, \citenamefont {Paramekanti},\ and\ \citenamefont
  {Kim}}]{BoYuan_PRB2017}%
  \BibitemOpen
  \bibfield  {author} {\bibinfo {author} {\bibfnamefont {B.}~\bibnamefont
  {Yuan}}, \bibinfo {author} {\bibfnamefont {J.~P.}\ \bibnamefont {Clancy}},
  \bibinfo {author} {\bibfnamefont {A.~M.}\ \bibnamefont {Cook}}, \bibinfo
  {author} {\bibfnamefont {C.~M.}\ \bibnamefont {Thompson}}, \bibinfo {author}
  {\bibfnamefont {J.}~\bibnamefont {Greedan}}, \bibinfo {author} {\bibfnamefont
  {G.}~\bibnamefont {Cao}}, \bibinfo {author} {\bibfnamefont {B.~C.}\
  \bibnamefont {Jeon}}, \bibinfo {author} {\bibfnamefont {T.~W.}\ \bibnamefont
  {Noh}}, \bibinfo {author} {\bibfnamefont {M.~H.}\ \bibnamefont {Upton}},
  \bibinfo {author} {\bibfnamefont {D.}~\bibnamefont {Casa}}, \bibinfo {author}
  {\bibfnamefont {T.}~\bibnamefont {Gog}}, \bibinfo {author} {\bibfnamefont
  {A.}~\bibnamefont {Paramekanti}}, \ and\ \bibinfo {author} {\bibfnamefont
  {Y.-J.}\ \bibnamefont {Kim}},\ }\href {\doibase 10.1103/PhysRevB.95.235114}
  {\bibfield  {journal} {\bibinfo  {journal} {Phys. Rev. B}\ }\textbf {\bibinfo
  {volume} {95}},\ \bibinfo {pages} {235114} (\bibinfo {year}
  {2017})}\BibitemShut {NoStop}%
\bibitem [{\citenamefont {Paramekanti}\ \emph {et~al.}(2018)\citenamefont
  {Paramekanti}, \citenamefont {Singh}, \citenamefont {Yuan}, \citenamefont
  {Casa}, \citenamefont {Said}, \citenamefont {Kim},\ and\ \citenamefont
  {Christianson}}]{Paramekanti_PRB2018}%
  \BibitemOpen
  \bibfield  {author} {\bibinfo {author} {\bibfnamefont {A.}~\bibnamefont
  {Paramekanti}}, \bibinfo {author} {\bibfnamefont {D.~J.}\ \bibnamefont
  {Singh}}, \bibinfo {author} {\bibfnamefont {B.}~\bibnamefont {Yuan}},
  \bibinfo {author} {\bibfnamefont {D.}~\bibnamefont {Casa}}, \bibinfo {author}
  {\bibfnamefont {A.}~\bibnamefont {Said}}, \bibinfo {author} {\bibfnamefont
  {Y.-J.}\ \bibnamefont {Kim}}, \ and\ \bibinfo {author} {\bibfnamefont
  {A.~D.}\ \bibnamefont {Christianson}},\ }\href {\doibase
  10.1103/PhysRevB.97.235119} {\bibfield  {journal} {\bibinfo  {journal} {Phys.
  Rev. B}\ }\textbf {\bibinfo {volume} {97}},\ \bibinfo {pages} {235119}
  (\bibinfo {year} {2018})}\BibitemShut {NoStop}%
\bibitem [{\citenamefont {Aharen}\ \emph {et~al.}(2010)\citenamefont {Aharen},
  \citenamefont {Greedan}, \citenamefont {Bridges}, \citenamefont {Aczel},
  \citenamefont {Rodriguez}, \citenamefont {MacDougall}, \citenamefont {Luke},
  \citenamefont {Michaelis}, \citenamefont {Kroeker}, \citenamefont {Wiebe},
  \citenamefont {Zhou},\ and\ \citenamefont {Cranswick}}]{Aharen_PRB2010}%
  \BibitemOpen
  \bibfield  {author} {\bibinfo {author} {\bibfnamefont {T.}~\bibnamefont
  {Aharen}}, \bibinfo {author} {\bibfnamefont {J.~E.}\ \bibnamefont {Greedan}},
  \bibinfo {author} {\bibfnamefont {C.~A.}\ \bibnamefont {Bridges}}, \bibinfo
  {author} {\bibfnamefont {A.~A.}\ \bibnamefont {Aczel}}, \bibinfo {author}
  {\bibfnamefont {J.}~\bibnamefont {Rodriguez}}, \bibinfo {author}
  {\bibfnamefont {G.}~\bibnamefont {MacDougall}}, \bibinfo {author}
  {\bibfnamefont {G.~M.}\ \bibnamefont {Luke}}, \bibinfo {author}
  {\bibfnamefont {V.~K.}\ \bibnamefont {Michaelis}}, \bibinfo {author}
  {\bibfnamefont {S.}~\bibnamefont {Kroeker}}, \bibinfo {author} {\bibfnamefont
  {C.~R.}\ \bibnamefont {Wiebe}}, \bibinfo {author} {\bibfnamefont
  {H.}~\bibnamefont {Zhou}}, \ and\ \bibinfo {author} {\bibfnamefont
  {L.~M.~D.}\ \bibnamefont {Cranswick}},\ }\href {\doibase
  10.1103/PhysRevB.81.064436} {\bibfield  {journal} {\bibinfo  {journal} {Phys.
  Rev. B}\ }\textbf {\bibinfo {volume} {81}},\ \bibinfo {pages} {064436}
  (\bibinfo {year} {2010})}\BibitemShut {NoStop}%
\bibitem [{\citenamefont {Georges}\ \emph {et~al.}(2013)\citenamefont
  {Georges}, \citenamefont {Medici},\ and\ \citenamefont
  {Mravlje}}]{Georges2013}%
  \BibitemOpen
  \bibfield  {author} {\bibinfo {author} {\bibfnamefont {A.}~\bibnamefont
  {Georges}}, \bibinfo {author} {\bibfnamefont {L.~d.}\ \bibnamefont {Medici}},
  \ and\ \bibinfo {author} {\bibfnamefont {J.}~\bibnamefont {Mravlje}},\ }\href
  {\doibase 10.1146/annurev-conmatphys-020911-125045} {\bibfield  {journal}
  {\bibinfo  {journal} {Annual Review of Condensed Matter Physics}\ }\textbf
  {\bibinfo {volume} {4}},\ \bibinfo {pages} {137} (\bibinfo {year}
  {2013})}\BibitemShut {NoStop}%
\bibitem [{\citenamefont {Simon}\ and\ \citenamefont
  {Varma}(2002)}]{Varma_PRL2002}%
  \BibitemOpen
  \bibfield  {author} {\bibinfo {author} {\bibfnamefont {M.~E.}\ \bibnamefont
  {Simon}}\ and\ \bibinfo {author} {\bibfnamefont {C.~M.}\ \bibnamefont
  {Varma}},\ }\href {\doibase 10.1103/PhysRevLett.89.247003} {\bibfield
  {journal} {\bibinfo  {journal} {Phys. Rev. Lett.}\ }\textbf {\bibinfo
  {volume} {89}},\ \bibinfo {pages} {247003} (\bibinfo {year}
  {2002})}\BibitemShut {NoStop}%
\bibitem [{\citenamefont {Chandra}\ \emph {et~al.}(2002)\citenamefont
  {Chandra}, \citenamefont {Coleman}, \citenamefont {Mydosh},\ and\
  \citenamefont {Tripathi}}]{Chandra_Nat2002}%
  \BibitemOpen
  \bibfield  {author} {\bibinfo {author} {\bibfnamefont {P.}~\bibnamefont
  {Chandra}}, \bibinfo {author} {\bibfnamefont {P.}~\bibnamefont {Coleman}},
  \bibinfo {author} {\bibfnamefont {J.~A.}\ \bibnamefont {Mydosh}}, \ and\
  \bibinfo {author} {\bibfnamefont {V.}~\bibnamefont {Tripathi}},\ }\href
  {\doibase 10.1038/nature00795} {\bibfield  {journal} {\bibinfo  {journal}
  {Nature}\ }\textbf {\bibinfo {volume} {417}},\ \bibinfo {pages} {831}
  (\bibinfo {year} {2002})}\BibitemShut {NoStop}%
\bibitem [{\citenamefont {Dawson}\ \emph {et~al.}(1988)\citenamefont {Dawson},
  \citenamefont {Tibbs}, \citenamefont {Weathersby}, \citenamefont {Boekema},\
  and\ \citenamefont {Chan}}]{Dawson1988}%
  \BibitemOpen
  \bibfield  {author} {\bibinfo {author} {\bibfnamefont {W.~K.}\ \bibnamefont
  {Dawson}}, \bibinfo {author} {\bibfnamefont {K.}~\bibnamefont {Tibbs}},
  \bibinfo {author} {\bibfnamefont {S.~P.}\ \bibnamefont {Weathersby}},
  \bibinfo {author} {\bibfnamefont {C.}~\bibnamefont {Boekema}}, \ and\
  \bibinfo {author} {\bibfnamefont {K.~B.}\ \bibnamefont {Chan}},\ }\href
  {\doibase 10.1063/1.342212} {\bibfield  {journal} {\bibinfo  {journal}
  {Journal of Applied Physics}\ }\textbf {\bibinfo {volume} {64}},\ \bibinfo
  {pages} {5809} (\bibinfo {year} {1988})}\BibitemShut {NoStop}%
\bibitem [{\citenamefont {Foronda}\ \emph {et~al.}(2015)\citenamefont
  {Foronda}, \citenamefont {Lang}, \citenamefont {M\"oller}, \citenamefont
  {Lancaster}, \citenamefont {Boothroyd}, \citenamefont {Pratt}, \citenamefont
  {Giblin}, \citenamefont {Prabhakaran},\ and\ \citenamefont
  {Blundell}}]{Blundell2015}%
  \BibitemOpen
  \bibfield  {author} {\bibinfo {author} {\bibfnamefont {F.~R.}\ \bibnamefont
  {Foronda}}, \bibinfo {author} {\bibfnamefont {F.}~\bibnamefont {Lang}},
  \bibinfo {author} {\bibfnamefont {J.~S.}\ \bibnamefont {M\"oller}}, \bibinfo
  {author} {\bibfnamefont {T.}~\bibnamefont {Lancaster}}, \bibinfo {author}
  {\bibfnamefont {A.~T.}\ \bibnamefont {Boothroyd}}, \bibinfo {author}
  {\bibfnamefont {F.~L.}\ \bibnamefont {Pratt}}, \bibinfo {author}
  {\bibfnamefont {S.~R.}\ \bibnamefont {Giblin}}, \bibinfo {author}
  {\bibfnamefont {D.}~\bibnamefont {Prabhakaran}}, \ and\ \bibinfo {author}
  {\bibfnamefont {S.~J.}\ \bibnamefont {Blundell}},\ }\href {\doibase
  10.1103/PhysRevLett.114.017602} {\bibfield  {journal} {\bibinfo  {journal}
  {Phys. Rev. Lett.}\ }\textbf {\bibinfo {volume} {114}},\ \bibinfo {pages}
  {017602} (\bibinfo {year} {2015})}\BibitemShut {NoStop}%
\bibitem [{\citenamefont {Patri}\ \emph {et~al.}(2020)\citenamefont {Patri},
  \citenamefont {Khait},\ and\ \citenamefont {Kim}}]{Patri2020NFL}%
  \BibitemOpen
  \bibfield  {author} {\bibinfo {author} {\bibfnamefont {A.~S.}\ \bibnamefont
  {Patri}}, \bibinfo {author} {\bibfnamefont {I.}~\bibnamefont {Khait}}, \ and\
  \bibinfo {author} {\bibfnamefont {Y.~B.}\ \bibnamefont {Kim}},\ }\href
  {\doibase 10.1103/PhysRevResearch.2.013257} {\bibfield  {journal} {\bibinfo
  {journal} {Phys. Rev. Research}\ }\textbf {\bibinfo {volume} {2}},\ \bibinfo
  {pages} {013257} (\bibinfo {year} {2020})}\BibitemShut {NoStop}%
\end{thebibliography}%

\appendix

\section{Orbital Wavefunctions and $\mathbf{L}$ Matrices}
The $d$ orbital basis is constructed out of the $l_z$ eigenstates of the angular momentum $l=2$ manifold, as 

\begin{align}
\begin{split}
\ket{yz} &= {i \over \sqrt{2}} \left( \ket{-1} + \ket{1} \right) \equiv \ket{1}_\alpha \\
\ket{xz} &= {1 \over \sqrt{2}} \left( \ket{-1} - \ket{1} \right) \equiv \ket{2}_\alpha \\
\ket{xy} &= {i \over \sqrt{2}} \left( \ket{-2} - \ket{2} \right) \equiv \ket{3}_\alpha \\
\ket{x^2 - y^2} &= {1 \over \sqrt{2}} \left( \ket{-2} + \ket{2} \right) \equiv \ket{4}_\alpha \\
\ket{3z^2 - r^2} &= \ket{0} \equiv \ket{5}_\alpha
\end{split}
\end{align}
where the states $\ket{m}$ refer to $\ket{l=2,m}$ and states with the subscript $\alpha$ indicate the orbital basis. Since this is the basis we will be working with in this paper, the $\alpha$ index will be dropped. The $\ket{m}$ states in position space can be represented using Spherical Harmonics (employing the Condon-Shortley phase), and the particular linear combinations above ensure that the orbital wavefunctions are real, giving the so-called Tesseral Harmonics. 
In this basis, the angular momentum matrices can be constructed as 
\begin{align}
\begin{split}
L_x &= 
\left( \begin{array}{c c c | c c }
0 & 0 & 0 & -i & -i\sqrt{3} \\
0 & 0 &  i & 0 & 0 \\
0 & -i & 0 & 0 & 0 \\
\hline
i & 0 & 0 & 0 & 0 \\
i\sqrt{3} & 0 & 0 & 0 & 0
\end{array} \right)
\\
L_y &= 
\left( \begin{array} {c c c | c c}
0 & 0 & -i & 0 & 0 \\
0 & 0 &  0 & -i & i\sqrt{3} \\
i & 0 & 0 & 0 & 0 \\
\hline
0 &i & 0 & 0 & 0 \\
0 & -i\sqrt{3} & 0 & 0 & 0
\end{array} \right)
\\
L_z &= 
\left( \begin{array}{c c c | c c}
0 & i & 0 & 0 & 0 \\
-i & 0 & 0 & 0 & 0 \\
0 & 0 & 0 & 2i & 0 \\
\hline
0 & 0 & -2i & 0 & 0 \\
0 & 0 & 0 & 0 & 0
\end{array} \right)
\end{split}
\end{align}
The top left blocks in the above matrices show the $t_{2g}$ subspace, and it is clear that the angular momentum is completely quenched in the $e_g$ subspace.

\section{Perturbation theory}
We carry out a perturbation theory study, using $H_{\text{CEF}}$ (Equation \ref{cfham}) as the unperturbed Hamiltonian and treating $J_H$ (interactions) and $\lambda$ (SOC)
as perturbations. Working in the two-electron basis $\ket{\alpha_1,s_1 ; \alpha_2, s_2} \equiv c^\dagger_{\alpha_1,s_1}c^\dagger_{\alpha_2,s_2}\ket{0}$, where the $\alpha$'s are orbital indices, and the $s$'s are spin indices, the unperturbed eigenspace consists of three energy levels, $\{ 0, V_C, 2V_C \}$, with degeneracies $\{ 15, 24, 6 \}$. These correspond to double occupancy within the $t_{2g}$ level, shared occupancy between the $t_{2g}$ and $e_g$ levels, and double occupancy in the $e_g$ level, respectively. 
The perturbations couple these different sectors. For instance, SOC can excite an electron from a $t_{2g}$ level into an $e_g$ level, across the gap $V_C$. Similarly, pair
hopping can hop a pair of electrons from a $t_{2g}$ level into an $e_g$ level, across an energy gap $2V_C$. Treating such terms within perturbation theory we find that
In order to project out the $e_g$ subspace, we treat all such mixing terms adding second-order and third-order perturbation effects, which leads to an effective $t_{2g}$
subspace Hamiltonian. At second order, we find that $U'-U$, pair hopping, and magnetic Hund's coupling are renormalized differently,
but in a way that does not break spherical symmetry, i.e. the renormalized Kanamori couplings obey \cite{Georges2013}
$U' - U = J_P + J_H$ (where $J_P$ and $J_H$ denote respectively the strength of the interorbital pair hopping and magnetic Hund's coupling). 
Diagonalizing the resulting effective Hamiltonian,
which sums the full Hamiltonian projected to $t_{2g}$ levels with the above perturbed interactions, we find that the ground state remains a five-fold degenerate $J=2$
mutiplet. However, at third order, we find new interactions that arise in the effective Hamiltonian in the $t_{2g}$ manifold which cannot be described as renormalizations of 
existing interactions; specifically, there are terms schematically given by
\be
\Delta H^{(3)}_{L,L'} \!\!=\!\! \sum_{H,H'} \!\! \frac{(H_{\rm SOC})_{L,H} (H_{\rm Hund})_{H,H'} (H_{\rm SOC})_{H',L'}}{V_C^2} \notag
\ee
where $L,H$ refer to low and high energy states with $L$ having both electrons in the $t_{2g}$ orbitals, and $H$ having one electron in $t_{2g}$ and the other in $e_g$.
This term leads to a splitting of the $J=2$ manifold into a low energy non-Kramers doublet and a high energy magnetic triplet, with the splitting emerging at 
${\cal O}(\lambda^2 J_H/V_C^2)$ at large $V_C$.

\section{Orbital currents and magnetic fields}
In order to study the impact of ferro-octupolar order in generating time-reversal breaking electronic currents and magnetic fields, we explicitly write out the orbital wavefunctions
in position space which enter the angular momentum states.
For this, we multiply the radial part of the hydrogen-like wavefunction with the Tesseral Harmonic of the orbital. We use the following form for the radial wavefunction:
\begin{equation}
R_{nl}(r) = N_{nl}~ \rho^{l}\!(r) e^{-\rho(r)/2} L_{n-l-1}^{2l+1}(\rho(r))
\end{equation} 
where $n$ is the principal quantum number, $l$ is the angular momentum quantum number, and $\rho(r) = 2r/na(r)$. $L_{n-l-1}^{2l+1}$ is the generalized Laguerre polynomial, and $N_{nl}$ is a normalization constant. $a(r)$ is a function which captures the screening by the inner electrons, which we call the ``effective" Bohr radius. The function must be chosen such that 
\begin{equation}
\lim_{r \rightarrow 0}a(r) = a_0/Z_0;~ \lim_{r \rightarrow \infty}a(r) = a_0/Z_\infty
\end{equation}
where $a_0$ is the (hydrogen) Bohr radius, $Z_0$ is the bare charge of the nucleus, and $Z_\infty$ is the effective charge that 
an electron sees at large distances. We propose the following simple form:
\begin{equation}
a(r) = {a_0 \over Z(r)};~ Z(r) = Z_\infty + (Z_0 - Z_\infty) e^{-r/r_0}
\end{equation}
with $r_0$ being a tuning parameter which determines how the effective charge falls off with distance. For instance, for an Os$^{6+}$ ion, $Z_0=76$ and $Z_\infty = 7$ (since
all electrons except the one $5d$ electron we focus on will
contribute to screening at large distances). A reasonable value for $r_0$ 
is that it is smaller than the ionic radius $\sim 70\si{\pico \meter}$; we thus consider $r_0 = 10$-$20 \si{\pico \meter}$.
If we are interested in $5d$ electrons, the radial wavefunction is of the form
\begin{equation}
\!\!\!\! R_{52}(r) \!=\! N_{52} \left( {2r \over 5a(r)} \right)^2 \! e^{-r/5a(r)} L_2^5\left( {2r \over 5a(r)} \right)
\end{equation}
The normalization constant $N_{52}$ depends on $r_0$. Hence the full wavefunction is given by 
$\psi_{nl\alpha} = R_{nl}(r)Y_{l\alpha}(\theta, \phi)$,
where $Y_{l\alpha}$ is the Tesseral Harmonic associated with the $\alpha$ orbital.
The current operator thus becomes
\begin{equation} \label{current}
\mathbf{J} =  {ie\hbar \over 2m} \sum_{\alpha,\beta} \sum_s \left( \psi_{nl\alpha} \pmb{\nabla} \psi_{nl\beta} - \psi_{nl\beta} \pmb{\nabla} \psi_{nl\alpha} \right) c^\dagger_{\alpha,s} c^\pdg_{\beta,s} 
\end{equation} 
All the spatial dependence of the current is encoded in the factor $\psi_{nl\alpha} \pmb{\nabla} \psi_{nl\beta} - \psi_{nl\beta} \pmb{\nabla} \psi_{nl\alpha} \equiv \pmb{\xi}_{\alpha\beta}(r,\theta,\phi)$.  Since the wavefunctions can be separated into the radial and angular components, i.e. $\psi_{nl\alpha} = R_{nl}(r)Y_{l\alpha}(\theta, \phi)$, this factor becomes
\begin{equation}
\pmb{\xi}_{\alpha \beta} = R_{nl}^2 \left( Y_{l\alpha} \pmb{\nabla} Y_{l\beta} - Y_{l\beta} \pmb{\nabla} Y_{l\alpha} \right)
\end{equation}
From the exact diagonalization, we can obtain the ground state of the system as some linear combination of our basis states. Let us call this ground state $\ket{\psi_g}$:
\begin{equation}
\ket{\psi_g} = \sum_\Omega a_\Omega \ket{\Omega}
\end{equation}
where $\ket{\Omega}$ refers to our basis states of the form $\ket{\alpha_1,s_1;\alpha_2,s_2}$. 
Since we are interested in the matrix elements of the current in Equation \eqref{current} in this state, we can recast the problem as 
\begin{equation}
\braket{\mathbf{J}} = {ie\hbar \over 2m} \sum_{\alpha,\beta} w_{\alpha\beta} \text{ } \pmb{\xi}_{\alpha\beta}
\end{equation}
where each factor $\pmb{\xi}_{\alpha\beta}$ is associated with a `weight' $w_{\alpha\beta}$, given by
\begin{equation}
w_{\alpha\beta} = \sum_{\Omega,\Omega'}a^*_\Omega a_{\Omega'}\sum_s \bra{\Omega}c^\dagger_{\alpha,s}c_{\beta,s} \ket{\Omega'}
\end{equation}
It can be seen that $w_{\alpha\beta} = w_{\beta\alpha}^*$. The Hermiticity of $\mathbf{J}$ constrains the weights to be purely imaginary. 
Once the expectation value of the current density is obtained, we use Eq.~\ref{mfield} to compute the magnetic field. 

\end{document}